\documentclass[a4paper,11pt]{article}
\usepackage[toc,page]{appendix}
\usepackage{jheppub}
\usepackage{setspace}
\usepackage{amssymb}
\usepackage{amsmath,amsfonts}
\usepackage{graphicx}
\usepackage{mathrsfs}
\usepackage[vcentermath]{youngtab}
\usepackage{color}
\numberwithin{equation}{section}
\usepackage{bbm}
\newcommand{\nn}{\nonumber\\}

\newcommand{\be}{\begin{equation}}
\newcommand{\ee}{\end{equation}}
\newcommand{\bea}{\begin{eqnarray}}
\newcommand{\eea}{\end{eqnarray}}

\newcommand{\eq}{&=&}
\newcommand{\comm}[2]{\left[ #1 \, , \, #2 \right]}
\newcommand{\acomm}[2]{\left\{ #1 \, , \, #2 \right\}}
\newcommand{\ket}[1]{\left\lvert #1 \right\rangle}
\newcommand{\lb}{\left(}
\newcommand{\rb}{\right)}
\usepackage{multirow}
\newcommand{\half}{\frac{1}{2}}

\newcommand{\third}{\frac{1}{3}}
\newcommand{\sixth}{\frac{1}{6}}

\newcommand{\ml}{\mathcal{L}}
\newcommand{\mk}{\mathcal{K}}

\renewcommand{\a}{\alpha}

\renewcommand{\b}{\beta}

\renewcommand{\d}{\delta}
\newcommand{\g}{\gamma}

\newcommand{\e}{{\epsilon }}

\newcommand{\s}{{\,\,\, }}
\newcommand{\mfg}{\mathfrak{g}}
\newcommand{\mcT}{\mathcal{T}}

\newcommand{\stln}{\setlength{\unitlength}{2.2ex}}
\newcommand{\fr}{\framebox(1,1){}}

\newcommand{\marcnbox}
{\stln \lower1.4ex \hbox{
\begin{picture} (6.6, 3.1)
\multiput(.3, .3) (1, 0) {2} {\fr}
\put(2.3, .3) {\framebox(3,1){$\cdots$}}
\put(5.3, .3){\fr}
\put(.3, 1.4)
{$\overbrace{~~~~~~~~~~~~~~~~~~}^{n}$}
\end{picture}}}

\newcommand{\bep}{\begin{picture}}
\newcommand{\eep}{\end{picture}}

\newcounter{YoungHeight}\newcounter{YoungWidth}

\newcounter{Mul1}\newcounter{Mul2}\newcounter{Mul3}\newcounter{Mul4}
\newcounter{A1}\newcounter{A2}
\newcounter{B3}
\newcounter{C3}\newcounter{C4}

\newcounter{T0}\newcounter{T1}

\newcounter{R0}

\newlength{\txtHShift}

\newlength{\txtWidth}

\newcommand{\Add}[3]{\setcounter{#1}{#2}\addtocounter{#1}{#3}}

\newcommand{\Length}[1]{#10}
\newcommand{\YoungScale}{}

\newcommand{\BlockA}[2]{{\YoungScale\bep(\Length{#1},\Length{#2}){\Add{A1}{#1}{1}\Add{A2}{#2}{1}}%
\multiput(0,0)(10,0){\value{A1}}{\line(0,1){\Length{#2}}}\multiput(0,0)(0,10){\value{A2}}{\line(1,0){\Length{#1}}}%
\setcounter{YoungHeight}{\Length{#2}}\setcounter{YoungWidth}{\Length{#1}}\eep}}

\newcommand{\YoungB}{\BlockA{2}{1}}

\newcommand{\YoungAA}{\BlockA{1}{2}}

\newcommand{\YoungBB}{\BlockA{2}{2}}

\newcommand{\YoungAAAA}{\BlockA{1}{4}}

\begin{document}
\begin{flushright}
 CERN-PH-TH/2014-011 \\
 AEI-2014-001
\end{flushright}

\title{Deformed Twistors and Higher Spin Conformal (Super-)Algebras in Six Dimensions}

\author{Karan Govil$^a$  and  } 
\author{Murat G\"unaydin$^{a,b,c}$}

\affiliation{$^a$ Institute for Gravitation and the Cosmos \\
 Physics Department ,
Pennsylvania State University\\
University Park, PA 16802, USA}
\affiliation{$^{b}$ Theory Division, Physics Department \\ CERN 
CH-1211 Geneva,  Switzerland  } 
\affiliation{$^{c}$ Max-Planck-Institut f\"{u}r Gravitationsphysik, \\
Albert-Einstein-Institut, \\
Am M\"{u}hlenberg 1, D-14476 Potsdam, Germany } 

\emailAdd{kzg126@psu.edu}
\emailAdd{murat@phys.psu.edu}
\abstract{Massless conformal scalar field in six dimensions  corresponds to the minimal unitary representation (minrep) of the conformal group $SO(6,2)$. This minrep admits  a family of ``deformations" labelled by the spin $t$ of an $SU(2)_T$ group, which is the $6d$ analog of helicity in four dimensions.  These deformations of the minrep of $SO(6,2)$  describe massless conformal fields that are symmetric tensors in the spinorial representation of the $6d$ Lorentz group. The minrep and its deformations were obtained by quantization of the nonlinear realization of $SO(6,2)$ as a quasiconformal group in arXiv:1005.3580. We give a novel reformulation of  the generators of $SO(6,2)$ for these representations as bilinears of \emph{deformed} twistorial oscillators which transform \emph{nonlinearly} under the Lorentz group $SO(5,1)$ and apply them to define higher spin algebras and superalgebras in $AdS_7$. The higher spin (HS)  algebra of Fradkin-Vasiliev type in $AdS_7$ is simply  the enveloping algebra of $SO(6,2)$ quotiented by a two-sided ideal (Joseph ideal) which annihilates the minrep. We show that the Joseph ideal vanishes identically for the quasiconformal realization  of the minrep and its enveloping algebra leads directly to the HS algebra in $AdS_7$. Furthermore, the enveloping algebras of the deformations of the minrep define a discrete infinite  family of HS algebras in $AdS_7$ for which certain $6d$ Lorentz covariant  deformations of the Joseph ideal vanish identically. These results extend to superconformal algebras $OSp(8^*|2N)$ and  we find a discrete infinite  family of HS superalgebras as enveloping algebras of the minimal unitary supermultiplet and its deformations.  Our results suggest the existence of a discrete  family of (supersymmetric) HS theories in $AdS_7$ which are dual to free (super)conformal field theories (CFTs) or to  interacting but integrable (supersymmetric) CFTs  in $6d$.
}

\maketitle
\section{Introduction} 
 The minimal unitary representation (minrep) of the $4d$ conformal group $SU(2,2)$ describes a massless scalar field  in four dimensions which admits a one-parameter family of deformations that describe massless fields of arbitrary helicity\cite{Fernando:2009fq}. The minrep and its deformations  were obtained  by quantization of the nonlinear realization of the four dimensional conformal  group $SU(2,2)$ as a quasiconformal group in a five dimensional space. Using the results of \cite{Fernando:2009fq} in a previous work \cite{Govil:2013uta} we showed 
 that the generators of $SU(2,2)$ for the minrep and its deformations  can be written as bilinears of \emph{deformed} twistorial oscillators which transform \emph{nonlinearly} under the Lorentz group $SL(2,\mathbb{C})$ and applied them to define and study higher spin algebras and superalgebras in $AdS_5$ whose isometry group is $SU(2,2)$\footnote{ Following Vasiliev we refer to them as $AdS_{d+1}/CFT_d$ higher spin algebras.}.  More specifically, 
the standard higher spin (HS)  algebra of Fradkin-Vasiliev type in $AdS_5$ is simply  the enveloping algebra of $SU(2,2)$ quotiented by a two-sided ideal 
(Joseph ideal) which annihilates the minrep. The Joseph ideal vanishes identically for the quasiconformal realization  of the minrep and hence its enveloping algebra leads directly to the standard HS algebra in $AdS_5$. 
The enveloping algebras of the deformed minreps define a one parameter family of HS algebras in $AdS_5$ for which certain $4d$ covariant  deformations of the Joseph ideal vanish identically in their quasiconformal realizations. These results extend fully to the quasiconformal realizations of the superconformal algebras $SU(2,2|N)$ with the even subgroup $SU(2,2) \times U(N)$ and  one obtains  a one parameter family of HS superalgebras as enveloping algebras of the minimal unitary representation and its deformations in their quasiconformal realizations.  
As we argued in  \cite{Govil:2013uta} these results suggest the existence of a  family of (supersymmetric) HS theories in $AdS_5$ which are dual to free (super)conformal field theories (CFTs) or to  interacting but integrable (supersymmetric) CFTs  in $4d$.  The corresponding picture for $AdS_4/CFT_3$ higher spin algebras is much simpler since  the minimal unitary representation of $SO(3,2)$ is the scalar singleton   of Dirac and it admits a unique ``deformation"   which is the spinor singleton whose oscillator realizations involve only bilinears and  hence the corresponding higher spin algebras  are simply the enveloping algebras    corresponding  to free field realizations \cite{Gunaydin:1989um}.

That the Fradkin-Vasiliev higher spin algebra in $AdS_4$ \cite{Fradkin:1986qy} corresponds simply to  the enveloping algebra of the singletonic realization of $Sp(4,\mathbb{R})$ was  pointed out in \cite{Gunaydin:1989um}\footnote{For related work see also \cite{Konshtein:1988yg}.}.  Again in \cite{Gunaydin:1989um} it was pointed out that the higher spin algebras in $AdS_5$ and $AdS_7$ and their supersymmetric extensions  could be similarly constructed as enveloping algebras of the doubletonic realizations of the super algebras $SU(2,2|N)$  and $OSp(8^*|2N)$.  A purely bosonic higher spin algebra in $AdS_7$ was studied along these lines in \cite{Sezgin:2001ij} using the doubletonic  realization of $SO(6,2)$  given in \cite{Gunaydin:1984wc,Fernando:2001ak}.  Higher spin superalgebras in dimensions $d > 3$ were also studied by Vasiliev in \cite{Vasiliev:2004cm}. 
However higher spin superalgebras in $AdS_d$ for $d>4$ studied by Vasiliev do not possess any finite  dimensional conventional supersymmetric subsuperalgebras. 
Cubic interactions for simple mixed-symmetry fields in HS theories in higher dimensional $AdS$ space times using Vasiliev's approach were investigated in \cite{Alkalaev:2010af,Boulanger:2011se}. In a subsequent paper it was claimed that the purely bosonic HS theory in $AdS_7$ is unique under certain  assumptions \cite{Boulanger:2013zza}.
 We refer  to \cite{Govil:2013uta} and the reviews \cite{Vasiliev:1995dn,Vasiliev:1999ba,Bekaert:2005vh,Iazeolla:2008bp,Sagnotti:2011qp,Didenko:2014dwa} for additional references on higher spin theories. 
 
  In this paper we extend the results of \cite{Govil:2013uta} to six dimensions and define and study $AdS_7/CFT_6$ higher spin algebras and superalgebras using the quasiconformal approach. 
Contrary to earlier claims in the literature we find a discrete infinite family of $AdS_7/CFT_6$ higher spin algebras and their supersymmetric extensions for any number of super symmetries. These HS algebras are obtained as the enveloping algebras of the minimal unitary representation of $SO(6,2)$ and a discrete infinite family  of ``deformations" thereof. 

Here we should explain why we use the term deformation in referring to a discretely labelled family  of HS algebras.  In an earlier work \cite{Govil:2013uta} we showed that there exists an infinite family of inequivalent higher spin algebras and superalgebras in four dimensions labelled by the helicity which uniquely characterizes the unitary representations of the conformal group $SU(2,2)$ corresponding to massless $4d$ conformal fields. In the quasiconformal approach helicity is simply the deformation parameter corresponding to the eigenvalue of the little group $U(1)$ of massless particles in four dimensions which takes on continuous values. The little group of massless particles in six dimensions is $SO(4)=SU(2)_T\times SU(2)_A$ and massless particles are labelled by the eigenvalues $(j_T,j_A)$ of the little group, i.e the labels belong to a discrete set. In fact for the conformally massless unitary representations eigenvalues are of the form $(j_T,0)$ or of the form $(0,j_A)$. Thus the unitary representations that are the six dimensional analogs of ``deformations" of the minrep in four dimensions are discretely labelled. Hence we refer to them as {\it discrete deformations} of the minrep. This discrete infinite family of HS algebras admit supersymmetric extensions for arbitrary number of supersymmetries. 

The plan of the paper is as follows:  We start by reviewing the covariant twistorial (doubleton) construction in section \ref{sect-twist} following \cite{Gunaydin:1984wc,Gunaydin:1999ci,Fernando:2001ak} and its reformulation in terms of Lorentz covariant twistorial oscillators \cite{Fernando:2001ak,Chiodaroli:2011pp}. Next in sections \ref{sect-qcg}-\ref{sect-qcgsusy} we present a novel reformulation of the quasiconformal realization of the minimal unitary representation (minrep) of $SO(6,2)$ and its supersymmetric extensions and their deformations \cite{Fernando:2010dp,Fernando:2010ia} in terms of deformed twistors that transform nonlinearly under the $6d$ Lorentz group. In section \ref{sect-joseph} we review the Eastwood's formula for the generators $\mathscr{J}$ of the annihilator of the minrep (Joseph ideal) and show by explicit calculations that it vanishes identically as an operator for the quasiconformal realization of the minrep. Then in section \ref{sect-joseph-deformations} we use the $6d$ Lorentz covariant formulation of the Joseph ideal to identify the deformed generators $\mathscr{J}_t$ that are the annihilators of the deformations of the minrep. These discrete deformations are labelled by the eigenvalues of an $SU(2)_G$ symmetry realized as bilinears of fermionic oscillators. Interestingly, the $6d$ analog of the Pauli-Lubansky vector does not vanish for the deformed minreps and becomes an (anti-)self-dual operator of rank three. Next we compare the generators of the Joseph ideal computed for doubleton realization and identify the analog of the deformation $SU(2)_G$ in the quasiconformal realization.  For the doubleton realization the generators of the Joseph ideal do not vanish as operators. They annihilate only the  subspace of the Fock space of the covariant oscillators corresponding to the minrep. In section \ref{sect-hsa} we define the $AdS_7/CFT_6$ higher spin algebra and its deformations as the enveloping algebra of the minrep and its deformations in the quasiconformal framework, respectively. We also extend  these results to corresponding higher spin superalgebras and  conclude with a brief discussion  in section \ref{sec-discussion}.

\section{Realizations of the $6d$ conformal algebra $SO(6,2)\sim SO^*(8)$ and its supersymmetric extension $OSp(8^*|4)$}
\label{sect-realizations}
\subsection{Covariant twistorial oscillator construction of the massless representations (doubletons) of $6d$ conformal group SO(6,2)}
\label{sect-twist}
In this subsection we shall review  the construction of  positive energy unitary representations of $SO(6,2)$ that  correspond to massless conformal fields in six dimensions following \cite{Gunaydin:1984wc,Gunaydin:1999ci,Fernando:2001ak}.
The Lie algebra of the conformal group $SO(6,2)$ in six dimensions is isomorphic to that of $SO^*(8)$ with the maximal compact subgroup $U(4)$. Commutation relations of the generators $M_{AB}$ of $SO(6,2)$ in the canonical basis have the form
\be
\comm{M_{AB}}{M_{CD}} = i (\eta_{BC}M_{AD}-\eta_{AC}M_{BD}-\eta_{BD}M_{AC}+\eta_{AD}M_{BC})
\ee
where $\eta_{AB} = \text{diag}(-,+,+,+,+,+,+,-)$ and $A,B = 0, \ldots, 7$.
Chiral spinor representation of $SO(6,2)$ can be written in terms in six-dimensional
gamma matrices $\Gamma_\mu$ that satisfy \[ \acomm{\Gamma_\mu}{\Gamma_\nu}
= -2\eta_{\mu\nu} \] where $\eta_{\mu\nu} = \text{diag}(-,+,+,+,+,+)$ and $\mu,\nu
= 0, \ldots, 5$ as follows\footnote{Opposite chirality spinor representation is obtained by taking $\Sigma_{\mu 7} := +\frac{1}{2}\Gamma_\mu$ and $\Sigma_{67}=+\frac{1}{2} \gamma_7$.}:
\be
\Sigma_{\mu\nu} := -\frac{i}{4}\comm{\Gamma_\mu}{\Gamma_\nu}, \qquad \Sigma_{\mu
6} := \frac{1}{2}\Gamma_\mu\Gamma_7, \qquad \Sigma_{\mu 7} := -\frac{1}{2}\Gamma_\mu,
\qquad \Sigma_{67} := -\frac{1}{2}\gamma_7
\ee
We adopt the conventions of \cite{Fernando:2001ak}  for gamma matrices:
\begin{align}
\Gamma_0 & = \sigma_3 \otimes \mathbbm{1}_2 \otimes \mathbbm{1}_2 \nn
\Gamma_1 & = i\sigma_1 \otimes \sigma_2 \otimes \mathbbm{1}_2 \nn
\Gamma_2 & = i\sigma_1 \otimes \sigma_1 \otimes \sigma_2 \nn
\Gamma_3 & = i\sigma_1 \otimes \sigma_3 \otimes \sigma_2 \nn
\Gamma_4 & = i\sigma_2 \otimes \mathbbm{1}_2 \otimes \sigma_2 \nn
\Gamma_5 & = i\sigma_2 \otimes \sigma_2 \otimes \sigma_1\nn
\Gamma_7 & = - \Gamma_0\Gamma_1\Gamma_2\Gamma_3\Gamma_4\Gamma_5
\end{align}
Consider the  bosonic oscillators $c_i,d_j$ and their hermitian conjugates $c^i,d^j$ respectively
$(i,j=1,2,3,4)$ that satisfy 
\be
\comm{c_i}{c^j}  = \delta_{i}^{j}, \qquad\qquad \comm{d_i}{d^j}
 = \delta_{i}^{j}
\ee
and  form a twistorial spinor operator $\Psi$ and its Dirac conjugate
$\overline{\Psi} = \Psi^\dagger \Gamma_0$ as :
\be
\Psi = \begin{pmatrix}
c_i \\ d^i \\
\end{pmatrix},
\qquad
\overline{\Psi} = \begin{pmatrix}
c^i, \,\,  - d_i
\end{pmatrix}
\ee
Then the bilinears $M_{AB} =\overline{\Psi} \Sigma_{AB} \Psi$ provide a realization of the Lie algebra of $SO(6,2)$:
\be
\comm{\overline{\Psi} \Sigma_{AB} \Psi}{\overline{\Psi} \Sigma_{CD}
\Psi} = \overline{\Psi} \comm{\Sigma_{AB}}{\Sigma_{CD}} \Psi
\ee
and the Fock space of the oscillators decompose into an infinite set of positive energy unitary irreducible representations of $SO(6,2)$ corresponding to  massless conformal fields in six dimensions. The resulting representations for one pair (color) of oscillators were called doubletons of $SO(6,2)$ in \cite{Gunaydin:1984wc,Gunaydin:1999ci,Fernando:2001ak}. 

The Lie algebra of the conformal group $SO(6,2)$ has a three-graded decomposition (referred to as compact three-grading) with respect to its maximal compact subalgebra ${\cal L}^{0} = SU(4) \times U(1)_{E}$,
\be
{SO}(6,2) = \mathcal{L}^{-} \oplus \mathcal{L}^{0} \oplus \mathcal{L}^{+},
\ee
where the three-grading is determined by the conformal Hamiltonian  $ E = \frac{1}{2} ( P_{0} + K_{0} )$. To construct positive energy unitary representations of $SO^{*}(8)$, one realizes the generators  as following billnears:
\bea
A_{ij}  &= & c_{i} d_{j} -  c_{j} d_{i} \, \in  \mathcal{L}^{-}, \qquad A^{ij}  =  c^i d^j - d^j d^i \, \in \, \mathcal{L}^{+} \nn
 M^{i}_{~j} &=&  c^i c_j + d_j d^i \, \in \mathcal{L}^{0}
\eea
where $i,j = 1,2,3,4$.

$M^{i}_{~j}$ generate the maximal compact subgroup $U(4)$.
The conformal Hamiltonian is given by the trace  $M^i{}_i$
\be
Q_{B} =  \frac{1}{2} M^{i}_{~i}  = \frac{1}{2} \left( N_{B} + 4  \right),
\ee
where $N_{B} \equiv c^{i}  c_{i} +  d^{i} d_{i}$ is the bosonic number operator. We shall denote the eigenvalues of $Q_{B}$ as $E$. The hermitian linear combinations of $A_{ij}$ and $A^{ij}$ are the non-compact generators of $SO(6,2)$ \cite{Gunaydin:1984wc,Gunaydin:1999ci,Fernando:2001ak}. Each lowest weight  (positive energy) UIR is uniquely determined by a set of states transforming in the lowest energy irreducible representation  $\ket{\Omega}$ of  $SU(4) \times U(1)_{E}$ that are annihilated by all the elements of $\mathcal{L}^{-}$ \cite{Gunaydin:1984wc,Gunaydin:1999ci}.\footnote{ Equivalently,  the lowest weight vector of the lowest energy irreducible representation of $SU(4)$ determines the UIR. Hence, by an abuse of terminology, we shall use interchangeably the terms ``lowest weight vector" and ``lowest energy irreducible representation". }
The possible lowest weight vectors for one pair of oscillators (doubletons)  in this compact basis  have $SU(4)$ Young tableaux with one row  \cite{Gunaydin:1999ci}, They are of the form
\bea
\ket{0}, & & \nn 
c^{i_1} \ket{0} & = & \ket{\yng(1)}, \nn 
c^{( i_1} c^{i_2 )} \ket{0} & = & \ket{\yng(2)}, \nn 
~ & \vdots & ~ \nn 
c^{( i_1} c^{i_2} \ldots c^{i_n )} \ket{0} & = & |{\marcnbox} \rangle, 
\eea 
plus
those obtained by interchanging $c$-type oscillators with $d$-type oscillators and  the
state 
\be c^{(i} d^{j)} \ket{0} = \ket{\yng(2)}.
\ee

The positive energy UIR's of $SO^*(8)$ can be identified with conformal fields in six dimensions transforming covariantly under the six-dimensional Lorentz group with definite conformal dimension. The Lorentz covariant spinorial oscillators are obtained from the $SU(4)$ covariant oscillators
by the action of an intertwining operator $T=e^{\frac{\pi}{2}M_{06}}$. We will use Greek letters for the Lorentz group
$SO(5,1) \sim SU^{*}(4)$ spinorial indices -- $\alpha,\beta = 1,2,3,4.$ Withour convention of gamma matrices, the Lorentz covariant
spinorial oscillators $\lambda_\a^{\s i}, \eta^{\a j}$ are given by: 
 \begin{align}
\lambda_\a^{\s 1} & =  \left( \begin{array}{c} c^{3} + d_1 \\ -c^{4} + d_2 \\ -c^{1} + d_3 \\ c^{2} + d_4 \end{array} \right),\qquad \lambda_\a^{\s 2}   =  \left( \begin{array}{c} -d^{3} + c_1 \\ d^{4} + c_2 \\ d^{1} + c_3 \\ -d^{2} + c_4 \end{array} \right) 
\end{align}
\begin{align}
\eta^{\a 1} & =  \left( \begin{array}{c} -c^{1} - d_3 \\ -c^{2} + d_4 \\ -c^{3} + d_1 \\ -c^{4} - d_2 \end{array} \right),\qquad  \eta^{\a 2} =  \left( \begin{array}{c} d^{1} - c_3 \\ d^{2} + c_4 \\ d^{3} + c_1 \\ d^{4} - c_2 \end{array} \right) \ . 
  \label{stuvspinors}
\end{align}
They satisfy the following commutation relations:
\be
\comm{\eta^{\a i}}{\lambda_\b^{\s j}} = -2\d_\b^\a \epsilon^{ij}
\ee
where $i,j = 1, 2 $ ($\epsilon_{12}=\epsilon^{21}=+1$) and $\a, \b  =1,2,3,4$ .
One finds that \cite{Chiodaroli:2011pp}
\be
\left(\Sigma^\mu P_{\mu} \right)_{\a\b}   = P_{\a\b} = \lambda_\a^{\s i} \lambda_\b^{\s j} \epsilon_{ij}
\quad,\qquad
\left(\bar{\Sigma}^{\mu} K_{\mu} \right)^{\a\b} = K^{\a\b}  = - \eta^{\a i } \eta^{\b j } \epsilon_{ij}
\ee
where  $\Sigma$-matrices in $d = 6$ are the analogs of Pauli matrices $\sigma_{\mu}$ in $d = 4$. The explicit form of $\Sigma^\mu, \bar{\Sigma}^\mu$ is given in Appendix \ref{sigmaold}. Note that the form of $\lambda_\a^{\s i}, \eta^{\a j}$ is slightly different from \cite{Fernando:2001ak,Chiodaroli:2011pp} as we are in the mostly positive signature and also the form of intertwining operator is slightly different. 

Similarly we can define Lorentz generators with spinor indices as follows:
\be
M_\a^{\s \b} = -\frac{i}{2}\lb \Sigma^\mu \bar{\Sigma}^\nu \rb_\a^{\s \b} M_{\mu\nu}
\ee
In terms of spinors $\lambda_\a^{\s i}, \eta^{\a j}$, they are given as follows:
\be
M_\a^{\s \b} = -\frac{1}{2}\lb \lambda_\a^{\s i} \eta^{\b j} - \frac{1}{4} \d^\b_\a \lambda_\g^{\s i} \eta^{\g j} \rb \epsilon_{ij}
\ee
The dilatation generator is given by:
\be
\Delta = \frac{i}{8} \lb \eta^{\a i} \lambda_\a^{\s j}  - \lambda_\a^{\s i} \eta^{\a j} \rb \epsilon_{ij}
\ee
The conformal algebra in terms of these generators is as follows:
\be
\comm{M_\a^{\s \b}}{M_\g^{\s \d}} = \d_\g^\b M_\a^{\s \d} - \d_\a^\d M_\g^{\s \b} 
\ee
\be
\comm{P_{\a\b}}{M_\g^{\s\d}} = 2 \d^\d_{[\a} P_{\b]\g} + \frac{1}{2} \d^\d_\g P_{\a\b}, \quad \comm{K^{\a\b}}{M_\g^{\s\d}} = -2 \d_\g^{[\a} K^{\b]\d} - \frac{1}{2} \d^\d_\g K^{\a\b}
\ee
\be
\comm{P_{\a\b}}{K^{\g\d}} = 16 \lb \d_{[\a}^{[\g} M_{\b]}^{\s \d]} - \frac{i}{2} \d_{[\a}^{[\g} \d_{\b]}^{\s \d]} \Delta \rb
\ee
\be
\comm{\Delta}{P_{\a\b}} = i P_{\a\b}, \qquad \comm{\Delta}{K^{\a\b}} = -i K^{\a\b}, \qquad\comm{\Delta}{M_\a^{\s \b}} = 0 
\ee
The doubletons correspond to massless conformal fields in six dimensions that transform as  symmetric tensors $\Psi_{\alpha \beta\gamma ....} \equiv \Psi_{(\alpha\beta\gamma...)}$ in their spinor indices

\subsection{Quasiconformal approach to minimal unitary representation of $SO(6,2)$ }
\label{sect-qcg}
The construction of the minimal unitary representation of the $6d$ conformal group $SO(6,2)$ by quantization of its quasiconformal action and its deformations were given in \cite{Fernando:2010dp,Fernando:2010ia}. In this section we will reformulate the generators of these representations in terms of deformed twistorial oscillators  as was done for $4d$ conformal group $SU(2,2)$ in \cite{Govil:2013uta}

The group $SO(6,2) \sim SO^*(8)$ can be realized as a quasiconformal group that leaves light-like separations with respect to a quartic distance function in nine dimensions invariant. The quantization of this geometric action leads to a nonlinear realization of the generators of $SO(6,2)$ in terms of a singlet coordinate  $x$, its conjugate momentum $p$ and four bosonic oscillators  $a_m,a^m$ and $b_m,b^m$, ($m.n=1,2$) satisfying\cite{Fernando:2010dp}:
\be
\comm{x}{p} =i, \qquad\qquad \comm{a_m}{a^n} =\d_m^n, \qquad\qquad \comm{b_m}{b^n} =\d_m^n
\ee
The semisimple component of the little group of massless particles in six dimensions is $SO(4)$ which can be written as $SU(2)_S\times SU(2)_A$. Its normalizer inside $SO(6,2)$ is $SO(2,2)$ which also decomposes as $SU(1,1)_K \times SU(1,1)_N$.  
The generators of  $SU(2)_S$ and of $SU(2)_A$ subgroups of $SO^*(8)$ are  realized as bilinears of $a$ and $b$ type oscillators within the quasiconformal approach as follows:
\bea
S_+ &= &a^m b_m, \qquad S_- = b^m a_m, \qquad S_0 = \half \lb N_a - N_b \rb \label{gen-su(2)S} \\
A_+ &=& a^1 a_2 + b^1 b_2, \quad A_- = a_1 a^2 + b_1 b^2, \quad A_0 = \half \lb a^1a_1 - a^2a_2 + b^1b_1 - b^2b_2 \rb
\eea
where $N_a = a^m a_m$ and $N_b =b^m b_m$ are the respective number operators. They satisfy 
\bea
\comm{S_0}{S_{\pm}} = \pm S_\pm, \qquad \comm{S_+}{S_-} = 2S_0 \\
\comm{A_0}{A_{\pm}} = \pm A_\pm, \qquad \comm{A_+}{A_-} = 2A_0 
\eea

In the previous section we followed conventions given in \cite{Fernando:2001ak} for the $6d$ sigma matrices ($\Sigma^\mu,\bar{\Sigma}^\mu$) within  the covariant twistorial construction of doubletons. In order to make  contact with the spinor-helicity formalism used in the amplitudes literature, we will follow the Clifford algebra conventions of \cite{Cheung:2009dc}, summarized  in Appendix \ref{app-conventions}, for the minimal unitary representation and its deformations within the quasiconformal approach. To avoid confusion, we will denote these  $6d$  sigma matrices as $\hat{\sigma}^\mu,\bar{\hat{\sigma}}^\mu$ ($\mu,\nu=0,1, \ldots 5$)

To express the momentum generators of $SO(6,2)$ of the minrep we shall introduce two sets of deformed twistors $Z_\a^{\s i}$ and $\widetilde{Z}_\a^{\s i}$ ($\a,\b = 1,2,3,4$, $i,j=1,2$) that transform nonlinearly under the Lorentz group (their commutation relations are given in Appendix \ref{app-YZcomm}):
\begin{align}
\label{Zdef1}
Z_1^{\s 1} &= b_1 - \half \lb x-ip \rb +\frac{1}{x} \lb S_0 + \frac{3}{4} \rb, & Z_1^{\s 2} &= a_1 - \frac{S_-}{x} \\
Z_2^{\s 1} &=  b_2 - \frac{S_+}{x}, & Z_2^{\s 2} &=  a_2 - \half \lb x-ip \rb -\frac{1}{x} \lb S_0 - \frac{3}{4} \rb \\
Z_3^{\s 1} &= -a^2 + \half \lb x+ip \rb +\frac{1}{x} \lb S_0 + \frac{3}{4} \rb, & Z_3^{\s 2} &= b^2 - \frac{S_-}{x}  \\
Z_4^{\s 1} &= a^1 - \frac{S_+}{x}, & Z_4^{\s 2} &= -b^1 + \half \lb x+ip \rb -\frac{1}{x} \lb S_0 - \frac{3}{4} \rb  
\label{Zdef2}
\end{align}
\begin{align}
\label{Ztdef1}
\widetilde{Z}_1^{\s 1} &= b_1 - \half \lb x-ip \rb +\frac{1}{x} \lb S_0 - \frac{3}{4} \rb, & \widetilde{Z}_1^{\s 2} &= a_1 - \frac{S_-}{x}  \\
\widetilde{Z}_2^{\s 1} &=  b_2 - \frac{S_+}{x} & \widetilde{Z}_2^{\s 2}, &=  a_2 - \half \lb x-ip \rb -\frac{1}{x} \lb S_0 + \frac{3}{4} \rb  \\
\widetilde{Z}_3^{\s 1} &= -a^2 + \half \lb x+ip \rb +\frac{1}{x} \lb S_0 - \frac{3}{4} \rb, & \widetilde{Z}_3^{\s 2} &= b^2 - \frac{S_-}{x}  \\
\widetilde{Z}_4^{\s 1} &= a^1 - \frac{S_+}{x}, & \widetilde{Z}_4^{\s 2} &= -b^1 + \half \lb x+ip \rb -\frac{1}{x} \lb S_0 + \frac{3}{4} \rb  
\label{Ztdef2}
\end{align}
In terms of these deformed twistors the momentum generators can then be written as bilinears:
\be
P_{\a\b}= Z_\a^{\s i} \widetilde{Z}_\b^{\s j} \e_{ij}
\ee
We should stress the fact that even though  the above deformed twistors transform nonlinearly under the Lorentz group the bilinears $P_{\a\b}$ transform covariantly as anti-symmetric tensors in spinorial indices. 

In order to realize the special conformal generators, we need another set of deformed twistors $Y^{\a i}$ and $\widetilde{Y}^{\a i}$ (their commutation relations are given in Appendix \ref{app-YZcomm}):
\begin{align}
\label{Ydef1}
Y^{11} &= a^1 + \frac{S_+}{x}, & Y^{12} &= -b^1 - \half \lb x+ip \rb +\frac{1}{x} \lb S_0 - \frac{3}{4} \rb \\
Y^{21} &= a^2 + \half \lb x+ip \rb +\frac{1}{x} \lb S_0 + \frac{3}{4} \rb, & Y^{22} &=  -b^2 - \frac{S_-}{x}  \\ 
Y^{31} &= -b_2 - \frac{S_+}{x}, & Y^{32} &= -a_2 - \half \lb x-ip \rb -\frac{1}{x} \lb S_0 - \frac{3}{4} \rb \\
Y^{41} &= b_1 + \half \lb x-ip \rb -\frac{1}{x} \lb S_0 + \frac{3}{4} \rb, & Y^{42} &=  a_1 + \frac{S_-}{x}
\label{Ydef2}
\end{align}
\begin{align}
\label{Ytdef1}
\widetilde{Y}^{11} &= a^1 + \frac{S_+}{x}, & \widetilde{Y}^{12} &= -b^1 - \half \lb x+ip \rb +\frac{1}{x} \lb S_0 + \frac{3}{4} \rb \\
\widetilde{Y}^{21} &= a^2 + \half \lb x+ip \rb +\frac{1}{x} \lb S_0 - \frac{3}{4} \rb, & \widetilde{Y}^{22} &=  -b^2 - \frac{S_-}{x}  \\
\widetilde{Y}^{31} &= -b_2 - \frac{S_+}{x}, & \widetilde{Y}^{32} &= -a_2 - \half \lb x-ip \rb -\frac{1}{x} \lb S_0 + \frac{3}{4} \rb \\
\widetilde{Y}^{41} &= b_1 + \half \lb x-ip \rb -\frac{1}{x} \lb S_0 - \frac{3}{4} \rb, & \widetilde{Y}^{42} &=  a_1 + \frac{S_-}{x}
\label{Ytdef2}
\end{align}
that transform nonlinearly under the Lorentz group.
The special conformal  generators can then be written as bilinears:
\be
K^{\a\b}= Y^{\a i} \widetilde{Y}^{\b j} \e_{ij}
\ee
which transform covariantly under the Lorentz group. 

The Lorentz subgroup $SO(5,1)\sim SU^*(4)\sim Sl(2,\mathbb{H})$  generators of the minrep of $SO(6,2)$ with spinorial  indices take the form
\be
M_\a^{\s \b} = -\frac{i}{2}\lb \hat{\sigma}^\mu \bar{\hat{\sigma}}^\nu \rb_\a^{\s \b} M_{\mu\nu}
\ee
which in terms of deformed twistorial oscillators  $Y,Z$  can be written as:
\bea
M_\a^{\s \b} \eq -\half \lb Z_\a^{\s i}\widetilde{Y}^{\b j} -\frac{1}{4} \delta^\a_\b  Z_\g^{\s i}\widetilde{Y}^{\g j} \rb \e_{ij} \\
\eq \half \lb Y^{\b i}\widetilde{Z}_\a^{\s j} - \frac{1}{4} \delta^\a_\b  Y^{\g i}\widetilde{Z}_\g^{\s j} \rb \e_{ij}
\eea
The dilatation generator $\Delta$ takes the form:
\be
\Delta = \frac{i}{8} \lb Z_\a^{\s i}\widetilde{Y}^{\a j} -  Y^{\a i}\widetilde{Z}_\a^{\s j}  \rb \e_{ij}
\ee
The commutation relations of the generators of the minrep of the conformal algebra $SO(6,2)$ given above are as follows:
\be
\comm{M_\a^{\s \b}}{M_\g^{\s \d}} = \d_\a^\d M_\g^{\s \b} - \d_\g^\b M_\a^{\s \d}
\ee
\be
\comm{P_{\a\b}}{M_\g^{\s\d}} = -2 \d^\d_{[\a} P_{\b]\g} - \frac{1}{2} \d^\d_\g P_{\a\b}, \quad \comm{K^{\a\b}}{M_\g^{\s\d}} = 2 \d_\g^{[\a} K^{\b]\d} + \frac{1}{2} \d^\d_\g K^{\a\b}
\ee
\be
\comm{P_{\a\b}}{K^{\g\d}} = 16 \lb \d_{[\a}^{[\g} M_{\b]}^{\s \d]} + \frac{i}{2} \d_{[\a}^{[\g} \d_{\b]}^{\s \d]} \Delta \rb
\ee
\be
\comm{\Delta}{P_{\a\b}} = i P_{\a\b}, \qquad \comm{\Delta}{K^{\a\b}} = -i K^{\a\b}, \qquad\comm{\Delta}{M_\a^{\s \b}} = 0 
\ee

The algebra $\mathfrak{so}(6, 2)$ can be given a 3-graded decomposition with respect to the conformal Hamiltonian, which is referred to as the compact 3-grading and the generators in this basis are reproduced in Appendix \ref{app-qcg3grading} following \cite{Fernando:2010dp}.

\subsection{Deformations of the minimal unitary representation of  $SO(6,2)$ }
\label{sect-qc-def}
It was shown in \cite{Fernando:2010dp} that the minrep of $SO^*(8)$ that was studied in previous section is simply isomorphic to the scalar doubleton representation that describes a conformal massless scalar field in six dimensions. However we have seen in section \ref{sect-twist} that $SO^*(8)$ admits infinitely many doubleton representations corresponding to massless conformal fields transforming as symmetric tensors in the spinorial indices.  As in  the case of $4d$ conformal group $SU(2,2)$, it was shown in \cite{Fernando:2010dp} that there exists a discrete infinity of deformations to the minrep of $SO(6,2)$ labeled by the spin $t$ of an $SU(2)$ symmetry group which is the $6d$ analog of helicity in $4d$.  Allowing this spin $t$ to take all possible values, one obtains a discretely infinite set of deformations of the minrep which are isomorphic to  the doubleton representations.  In this section we will show that the generators of these representations  can be recast as bilinears of  deformed twistorial operators as was done  in the previous subsection for the true minrep that corresponds to $t=0$. 

Following \cite{Fernando:2010dp}, let us  introduce an arbitrary number $P$ pairs of fermionic oscillators $\rho_x$ and $\chi_x$ and their hermitian conjugates $\rho^x = (\rho_x)^\dagger$ and $\chi^x = (\chi_x)^\dagger$, ($x=1,2,\ldots,P$) that satisfy the anti-commutation relations:
\be
\acomm{\rho_x}{\rho^y} =\acomm{\chi_x}{\chi^y} = \d_x^y, \qquad \acomm{\rho_x}{\rho_y} =\acomm{\rho_x}{\chi_y} = \acomm{\chi_x}{\chi_y} = 0
\ee
and refer to them  as ``deformation fermions ".   The following bilinears of these fermionic oscillators
\begin{equation}
G_+ = \rho^x \chi_x
\qquad \qquad
G_- = \chi^x \rho_x
\qquad \qquad
G_0 = \frac{1}{2} \left( N_\rho - N_\chi \right)
\label{su2g}
\end{equation}
,where $N_\rho = \rho^x \rho_x$ and $N_\chi = \chi^x \chi_x$ are the respective
number operators, generate an $\mathfrak{su}(2)_G$ algebra:
\begin{equation}
\comm{G_+}{G_-} = 2 \, G_0, \qquad  \comm{G_0}{G_\pm} = \pm G_0
\end{equation}
The fermionic oscillators $\rho^x$ and $\chi^x$ form a doublet of $SU(2)_G$. 
We choose the Fock vacuum of these fermionic oscillators such that
\begin{equation}
\rho_x \ket{0} = \chi_x \ket{0} = 0
\end{equation}
for all $x = 1,2,\dots,P$. The states of the form
\begin{equation*}
\chi^{[x_1} \chi^{x_2} \dots \rho^{x_{(n-1)} } \rho^{{x_n}]} \ket{0}
\end{equation*}
with definite eigenvalue $n \leq P$ of the total number operator $N_T=N_\chi + N_\rho$ transform irreducibly in the spin $j = n/2$ representation of $SU(2)_G$ \footnote{
Note that square bracketing of  indices implies complete 
anti-symmetrization of weight one.}.  Among the irreducible representations of $SU(2)_G$ in the Fock space  of $P$ pairs of deformation fermions the multiplicity of the highest spin ( $j=P/2$ ) representation is one. 

Now to deform the minimal unitary 
realization of $\mathfrak{so}^*(8)$, one extends the subalgebra 
$\mathfrak{su}(2)_S$ to the diagonal subalgebra 
$\mathfrak{su}(2)_T$ of $\mathfrak{su}(2)_S$ and
$\mathfrak{su}(2)_G$ \cite{Fernando:2010dp}. In other words, the generators of $\mathfrak{su}(2)_S$ 
receive contributions from the $\rho$- and $\chi$-type fermionic oscillators 
as follows:
\begin{equation}
\begin{split}
T_+
&= S_+ + G_+
 = a^m b_m + \rho^x \chi_x
\\
T_-
&= S_- + G_-
 = b^m a_m + \chi^x \rho_x
\\
T_0
&= S_0 + G_0
 = \frac{1}{2} \left( N_a - N_b + N_\rho - N_\chi \right)
\end{split}
\label{gen-su(2)T}
\end{equation}
The quadratic Casimir of this subalgebra
$\mathfrak{su}(2)_T$ is given by
\begin{equation}
\mathcal{C}_2 \left[ \mathfrak{su}(2)_{T} \right]
= \,\, T^2
= \,\, T_0 T_0
  + \frac{1}{2}
    \left( \, T_+ T_-
           + T_- T_+
    \right) \,.
\end{equation}
In order to obtain the generators for the deformations of the minrep, all we need to do is replace $S_0, S_\pm$ by $T_0,T_\pm$ respectively in equations \ref{Zdef1} - \ref{Ztdef2} and equations \ref{Ydef1} - \ref{Ytdef2}. We will denote the resulting deformed twistors as $(Z_t)_\a^{\s i}, (\widetilde{Z_t})_\a^{\s i}$ and $(Y_t)^{\a i},(\widetilde{Y_t})^{\a i}$. The generators can then be written as:
\be
P_{\a\b}= (Z_t)_\a^{\s i} (\widetilde{Z_t})_\b^{\s j} \e_{ij}, \qquad K^{\a\b}= (Y_t)^{\a i} (\widetilde{Y_t})^{\b j} \e_{ij}
\ee
\bea
M_\a^{\s \b} \eq -\half \lb (Z_t)_\a^{\s i}(\widetilde{Y_t})^{\b j} -\frac{1}{4} \delta^\a_\b  (Z_t)_\g^{\s i}(\widetilde{Y_t})^{\g j} \rb \e_{ij} \\
\eq \half \lb (Y_t)^{\b i}(\widetilde{Z_t})_\a^{\s j} - \frac{1}{4} \delta^\a_\b  (Y_t)^{\g i}(\widetilde{Z_t})_\g^{\s j} \rb \e_{ij}
\eea
\be
\Delta = \frac{i}{8} \lb (Z_t)_\a^{\s i}(\widetilde{Y_t})^{\a j} -  (Y_t)^{\a i}(\widetilde{Z_t})_\a^{\s j}  \rb \e_{ij}
\ee
The Casimir invariants for $SO(6,2)$ for the deformed minreps depend only on the quadratic Casimir of $SU(2)_G$ involving deformation fermions. For one set of fermions ($P=1$)  one finds:
\bea
\label{cas1}
C_2 \eq M^A_{~B}M^B_{~A} = 16 - 6 \lb N_\rho -N_\chi \rb^2 \\
C_4 \eq M^A_{~B}M^B_{~C}M^C_{~D}M^D_{~A} = -\frac{69}{12}C_2 - 20 \\
C'_4 \eq \epsilon^{ABCDEFGH}M_{AB}M_{CD}M_{EF}M_{GH} = 60 \lb C_2-16 \rb \\
C_6 \eq M^A_{~B}M^B_{~C}M^C_{~D}M^D_{~E}M^E_{~F}M^F_{~A} = \frac{63}{48}C_2 + 475 
\eea
which shows clearly that the deformations are driven by fermionic oscillators.

\subsection{Minimal unitary supermultiplet of $OSp(8^*|4)$ and its deformations}
\label{sect-qcgsusy}
The construction of the minimal unitary representations of noncompact Lie algebras  by quantization of their quasiconformal realizations extends to noncompact Lie superalgebras \cite{Gunaydin:2006vz,Fernando:2009fq,Fernando:2010dp,Fernando:2010ia}. In this section we will reformulate the minimal unitary realization of   $6d$ superconformal algebra $OSp(8^*|4)$ with the even subgroup $SO^*(8) \times USp(4)$ given in \cite{Fernando:2010ia} in terms of deformed twistors. Extension to general superalgebras $OSp(8^*|2N)$ is straightforward.

Consider the superconformal (non-compact) 5-graded decomposition of the Lie superalgebra $\mathfrak{osp}(8^*|4)$ with respect to the dilatation generator $\Delta$:
\bea
\mathfrak{osp}(8^*|4) \eq \mathfrak{N}^{-1} \oplus \mathfrak{N}^{-1/2} \oplus \mathfrak{N}^0  \oplus \mathfrak{N}^{+1/2} \oplus \mathfrak{N}^{+1} \\ 
\eq K^{\a\b} \oplus S^{\a a} \oplus (M_\a^{\s\b} \oplus \Delta \oplus U^{ab}) \oplus Q_\a^{\s a} \oplus P_{\a\b}, \qquad (a,b  =1,2,3,4) \nonumber
\eea
where the grade zero space consists of the Lorentz algebra $\mathfrak{so}(5,1)$ ($M_\a^{\s\b}$), the dilatations ($\Delta$) and R-symmetry algebra $\mathfrak{usp}(4)$ ($U^{ab}$), grade $+1$ and $-1$ spaces consist of translations ($P_{\a\b}$) and special conformal transformations ($K^{\a\b}$) respectively, and the +1/2 and -1/2 spaces consist of Poincar\'e supersymmetries ($Q_\a^{\s a}$) and conformal supersymmetries ($S^{\a a}$) respectively. 

We introduce fermionic oscillators $\xi^{ai}$ where $a,b=1,2,3,4$ are the $USp(4) \sim SO(5)$ indices which are raised and lowered by the antisymmetric symplectic metric \[ \Omega^{ab}=\begin{pmatrix}
0 & \mathbbm{1}_2 \\
-\mathbbm{1}_2 & 0
\end{pmatrix}\]
and $i,j=1,2$ are the $SU(2)$ indices raised and lowered by $\e_{ij}$ ($\e_{12}=\e^{21}=+1$). These oscillators satisfy:
\be
\acomm{\xi^{ai}}{\xi^{bj}} = \Omega^{ab} \e^{ij}
\ee
and they will be referred to as  supersymmetry fermions. The following bilinears of these fermions
\be
F_+ = \half \xi^{a1}\xi^{b1} \Omega_{ba}, \qquad F_- = \half \xi^{a2}\xi^{b2} \Omega_{ab}, \qquad F_0 = \frac{1}{4} \lb \xi^{a1}\xi^{b2} + \xi^{a2}\xi^{b1} \rb \Omega_{ba}
\ee
generate a $\mathfrak{su}(2)_F$ algebra:
\begin{equation}
\comm{F_+}{F_-} = 2 \, F_0, \qquad \comm{F_0}{F_\pm} = \pm F_0
\end{equation}
To obtain the supersymmetric extensions of the deformations of the minrep of $SO^*(8)$, one extends the $\mathfrak{su}(2)_T$ subalgebra to $\mathfrak{su}(2)_\mcT$ which is the diagonal subalgebra of $\mathfrak{su}(2)_T$ and $\mathfrak{su}(2)_F$. The  generators of $\mathfrak{su}(2)_\mcT$ are then given by:
\bea
\mcT_+ \eq S_+ + G_+ + F_+ = a^i b_i + \rho^x \chi_x + \half \xi^{a1}\xi^{b1} \Omega_{ba} \\
\mcT_- \eq S_- + G_- + F_- = b^i a_i + \chi^x \rho_x + \half \xi^{a2}\xi^{b2} \Omega_{ab} \\
\mcT_0 \eq S_0 + G_0 + F_0 = \half (N_a - N_b + N_\rho - N_\chi) + \frac{1}{4} \lb \xi^{a1}\xi^{b2} + \xi^{a2}\xi^{b1} \rb \Omega_{ba}
\label{gen-su(2)mct}
\eea

The generators of the even subgroup $SO^*(8)$ of the deformations of the minrep of $OSp(8^*|4)$ are then obtained simply by replacing  $S_0, S_\pm$ by $\mcT_0,\mcT_\pm$ respectively in equations \ref{Zdef1} - \ref{Ztdef2} and equations \ref{Ydef1} - \ref{Ytdef2}. We will denote the resulting deformed twistors as $(Z^s_t)_\a^{\s i}, (\widetilde{Z^s_t})_\a^{\s i}$ and $(Y^s_t)^{\a i},(\widetilde{Y^s_t})^{\a i}$.
The generators of $SO^*(8)$  can then be written as bilinears of these deformed twistors:
\be
P_{\a\b}= (Z^s_t)_\a^{\s i} (\widetilde{Z^s_t})_\b^{\s j} \e_{ij}, \qquad K^{\a\b}= (Y^s_t)^{\a i} (\widetilde{Y^s_t})^{\b j} \e_{ij}
\ee
\bea
M_\a^{\s \b} \eq -\half \lb (Z^s_t)_\a^{\s i}(\widetilde{Y^s_t})^{\b j} -\frac{1}{4} \delta^\a_\b  (Z^s_t)_\g^{\s i}(\widetilde{Y^s_t})^{\g j} \rb \e_{ij} \\
\eq \half \lb (Y^s_t)^{\b i}(\widetilde{Z^s_t})_\a^{\s j} - \frac{1}{4} \delta^\a_\b  (Y^s_t)^{\g i}(\widetilde{Z^s_t})_\g^{\s j} \rb \e_{ij}
\eea
\be
\Delta = \frac{i}{8} \lb (Z^s_t)_\a^{\s i}(\widetilde{Y^s_t})^{\a j} -  (Y^s_t)^{\a i}(\widetilde{Z^s_t})_\a^{\s j}  \rb \e_{ij}
\ee
The supersymmetry generators $Q_\a^{\s a}$, $S^\a_{\s a}$ can similarly  be realized simply as bilinears of ordinary fermionic oscillators and deformed twistors as follows\footnote{To obtain the generators for the true minimal unitary supermultiplet
of $OSp(8^*|4)$ one  needs only to drop the deformation fermions from the generators of $\mathfrak{su}(2)_\mcT$ and the corresponding deformed twistors will be denoted as $(Z^s)_\a^{\s i}, (\widetilde{Z^s})_\a^{\s i}$ and $(Y^s)^{\a i},(\widetilde{Y^s})^{\a i}$.}:
\be
Q_\a^{\s a} = (Z^s_t)_\a^{\sp i} \xi^{aj} \e_{ij} = \xi^{ai} (\widetilde{Z^s_t})_\a^{\sp j} \e_{ij}
\ee
\be
S^{\a a} = (Y^s_t)^{\a i} \xi^{aj} \e_{ij} =  \xi^{ai} (\widetilde{Y^s_t})^{\a j} \e_{ij}
\ee
The supersymmetry generators satisfy
\be
\acomm{Q_\a^{\s a}}{Q_\b^{\s b}} = -\Omega^{ab} P_{\a\b}, \qquad \acomm{S^{\a a}}{S^{\b b}} = -\Omega^{ab} K^{\a\b}
\ee
\be
\acomm{S^{\a a}}{Q_\b^{\s b}} = -2 \Omega^{ab} M_\b^{\s \a} - i \delta^\a_\b \Omega^{ab} \Delta - 2  \delta^\a_\b U^{ab}
\ee
The $R$-symmetry group $USp(4)\sim SO(5)$ can realized as bilinears of fermionic oscillators as follows:
\be
U^{ab} = \lb \xi^{ai}\xi^{bj} - \frac{1}{4}\Omega^{ab} \Omega_{cd} \xi^{ci}\xi^{dj} \rb \e_{ij}
\ee
They satisfy the following commutation relations:
\be
\comm{U^{ab}}{U^{cd}} = 2 \Omega^{a(c} U^{d)b} + 2 \Omega^{b(c} U^{d)a}  
\ee
The commutators of $SO^*(8)$ generators  with the supersymmetry generators are as follows:
\begin{align}
\comm{M_\a^{\s \b}}{Q_\g^{\s a}} = - \d^\b_\g Q_\a^{\s a} + \frac{1}{4}\d^\b_\a Q_\g^{\s a},& \qquad \comm{M_\a^{\s \b}}{S^{\g a}} = \d^\g_\a S^{\b a} - \frac{1}{4}\d^\b_\a S^{\g a}  \\
\comm{K^{\a\b}}{Q_\g^{\s a}} = -4 \d_\g^{[\a}S^{\b]a},& \qquad \comm{P_{\a\b}}{S^{\g a}} = -4 \d^\g_{[\a}Q_{\b]}^{\s a} \\
\comm{\Delta}{Q_\g^{\a a}}  = \frac{i}{2}Q_\a^{\s a},& \qquad \comm{\Delta}{S^{\a a}} = -\frac{i}{2} S^{\a a}
\end{align}
The $R$-symmetry generators act on $USp(4)$ indices of supersymmetry generators as follows:
\be
\comm{U^{ab}}{Q_\a^{\s c}} = -2 \Omega^{c(a}Q_\g^{\s b)}, \qquad \comm{U^{ab}}{S^{\a c}} = -2 \Omega^{c(a}S^{\g b)}
\ee

The generators given above transform covariantly with respect to the  subgroup $SU^*(4)\times SO(1,1) \times  USp(4)$. 
Unitarity and positive energy nature of the resulting representations are made manifest by going to the compact three grading of $OSp(8^*|4)$ with respect to the compact sub-superalgebra $SU(4|2)\times U(1)$  \cite{Fernando:2010dp}.
\section{ $AdS_7/CFT_6$ higher spin (super-)algebras, Joseph ideals and their deformations}
\label{sect-joseph}

As reviewed in \cite{Govil:2013uta} the standard $AdS_{d}/CFT_{(d-1)} $  higher spin algebra of Fradkin-Vasiliev type is simply given by the quotient of the universal enveloping algebra of $SO(d,2)$   by a two-sided ideal \cite{Gunaydin:1989um,Vasiliev:1999ba,Eastwood:2002su,eastwood2005uniqueness}. This two-sided ideal is the Joseph ideal that annihilates the minimal unitary representation. 
Denoting the higher spin algebra as $HS(\mfg)$  with $\mfg =\mathfrak{so}(d-1,2)$   and the universal enveloping algebra as  $\mathscr{U}(\mfg)$ we have 
\be
HS(\mfg) = \frac{\mathscr{U}(\mfg)}{\mathscr{J}(\mfg)} 
\ee
where $\mathscr{J}(\mfg)$ denotes the Joseph ideal. 

The uniqueness of the Joseph ideal was proved in \cite{eastwood2005uniqueness} and an explicit formula for the generators of this ideal for $SO(n-2,2)$ was given as :
\bea
J_{ABCD} &=& M_{AB}M_{CD} - M_{AB}\circledcirc M_{CD} - \frac{1}{2}\comm{M_{AB}}{M_{CD}} + \frac{n-4}{4(n-1)(n-2)} \langle M_{AB},M_{CD}  \rangle \, \mathbf{1} \nonumber  \\
&=& \frac{1}{2}  M_{AB} \cdot M_{CD} - M_{AB}\circledcirc M_{CD} + \frac{n-4}{4(n-1)(n-2)} \langle M_{AB},M_{CD} \rangle \,  \mathbf{1} \label{Joseph} 
\eea
where the dot  $
\cdot$ denotes the symmetric product  \be M_{AB}\cdot M_{CD} \equiv M_{AB} M_{CD}+M_{CD}M_{AB} \ee
  $\langle M_{AB},M_{CD} \rangle$ is the Killing form of $SO(n-2,2)$ given by
\be
\langle M_{AB},M_{CD} \rangle = h\, M_{EF}M_{GH} (\eta^{EG}\eta^{FH}-\eta^{EH}\eta^{FG}) (\eta_{AC}\eta_{BD}-\eta_{AD}\eta_{BC})
\ee
where $h = \frac{2(n-2)}{n(4-n)}$ chosen such  that all possible contractions of $J_{ABCD}$ with the the metric vanish. The symbol $\circledcirc$ denotes the Cartan product of two generators \cite{eastwood2005cartan}:
\bea
M_{AB}\circledcirc M_{CD} \eq \third M_{AB}M_{CD} + \third M_{DC}M_{BA} + \sixth M_{AC}M_{BD} \nn
&& -\sixth M_{AD}M_{BC} + \sixth M_{DB}M_{CA} - \sixth M_{CB}M_{DA} \nn
&& -\frac{1}{2(n-2)}\left(M_{AE}M_C^E\eta_{BD}-M_{BE}M_C^E\eta_{AD}+M_{BE}M_D^E\eta_{AC}-M_{AE}M_D^E\eta_{BC}\right) \nn
&& -\frac{1}{2(n-2)}\left(M_{CE}M_A^E\eta_{BD}-M_{CE}M_B^E\eta_{AD}+M_{DE}M_B^E\eta_{AC}-M_{DE}M_A^E\delta_{BC}\right) \nn
&& +\frac{1}{(n-1)(n-2)} M_{EF}M^{EF}\left(\eta_{AC}\eta_{BD}-\eta_{BC}\eta_{AD}\right)
\eea
  We shall refer to the operator $J_{ABCD}$ as the generator of the Joseph ideal which for $SO(6,2)$ takes the form:
\be
J_{ABCD} =\frac{1}{2} M_{AB}\cdot M_{CD} - M_{AB}\circledcirc M_{CD} 
- \frac{1}{112} \langle M_{AB},M_{CD} \rangle
\label{joseph-6d}
\ee

The enveloping algebra $\mathscr{U}(\mathfrak{g})$ of a Lie algebra $\mathfrak{g}$ can be decomposed with respect to the  adjoint action of $\mathfrak{g}$. By Poincare-Birkhoff-Witt theorem this is equivalent to  computing symmetric products of the generators $M_{AB}\sim\parbox{10pt}{\YoungAA}\,$ of $\mathfrak{g}$ . 
For $ \mathfrak{so}(6,2)$ the symmetric product of the adjoint action decomposes as:
\be
\parbox{10pt}{\YoungAA} \otimes \parbox{10pt}{\YoungAA} = \parbox{20pt}{\YoungBB}\oplus\parbox{10pt}
{\YoungAAAA}\oplus\parbox{20pt}{\YoungB}\oplus\bullet
\label{ytdecomp}
\ee
where the singlet $\bullet$ is the quadratic Casimir $C_2 \sim M^{~A}_B M^{~B}_A$. It was pointed out in \cite{Vasiliev:1999ba} that the higher spin algebra $HS(\mfg)$ must be a quotient of $\mathscr{U}(\mfg)$ since the higher spin fields are described by traceless two row Young tableaux. Thus the relevant ideal should quotient out the all the diagrams except the first one in the above decomposition. The Joseph ideal as defined in \ref{Joseph} includes all the diagrams in the decomposition except the ``window" diagram $\parbox{20pt}{\YoungBB}$ and thus by quotienting $\mathscr{U}(\mathfrak{g})$ by the ideal generated by $J_{ABCD}$  defined in \ref{Joseph}, we get rid of all the ``unwanted" diagrams and obtain the Fradkin-Vasiliev type higher spin algebra $HS(6,2)$.

\subsection{Joseph ideal for minimal unitary representation of $SO(6,2)$}
Since the Eastwood formula for Joseph ideal is in the canonical basis we define 
\bea
M_{\mu 6} \eq \frac{1}{2} \lb P_\mu - K_\mu \rb \\
M_{\mu 7} \eq \frac{1}{2} \lb P_\mu + K_\mu \rb \\
M_{67} \eq -\Delta 
\eea
which together with the Lorentz group generators $M_{\mu\nu}$ form the canonical basis  $M_{AB}$ ($A,B,..=0,1,...7$). 
Substituting the expressions for the generators $M_{AB}$ of  the minrep  of $SO(6,2)$ from the quasiconformal  realization into the generator of Joseph ideal (equation \ref{joseph-6d}) one finds that {\it it vanishes identically as an operator}. 
To get a better insight into the physical meaning of the vanishing of the Joseph ideal  we write $J_{ABCD}$   in the Lorentz covariant conformal basis ($K_\mu, M_{\mu\nu}, \Delta, P_\mu$).which is equivalent to  certain quadratic identities.  In addition to the  conditions:
\be
P^\mu P_\mu = K^\mu K_\mu =0 \quad \text{or,} \quad P^2 = K^2 =0
\ee
one finds the  following identities:
\bea
6\Delta \cdot \Delta +  M^{\mu\nu} \cdot M_{\mu\nu} + 2  P^\mu \cdot K_\mu  \eq 0 \\
P^\mu \cdot (M_{\mu\nu} + \eta_{\mu\nu} \Delta) \eq 0 \\
K^\mu \cdot (M_{\nu\mu} + \eta_{\nu\mu} \Delta) \eq 0
\eea

\bea
 \eta^{\mu\nu} M_{\mu\rho} \cdot M_{\nu\sigma} - P_{(\rho} \cdot K_{\sigma)}  + 4 \eta_{\rho\sigma} \eq 0 \label{mpident6}\\
M_{\mu\nu} \cdot M_{\rho\sigma} + M_{\mu\sigma}\cdot M_{\nu\rho} + M_{\mu\rho}\cdot M_{\sigma\nu}  \eq 0  \label{mident6}\\
\Delta \cdot M_{\mu\nu} + P_{[\mu} \cdot K_{\nu]} \eq 0 \\
M_{[\mu\nu} \cdot P_{\rho]} \eq 0 \\ M_{[\mu\nu} \cdot K_{\rho]} \eq 0
\label{dmident6}
\eea
Defining the generalized Pauli-Lubanski tensor and its conformal analogue in six dimensions as:
\be
A_{\mu\nu\rho} =\frac{1}{3!}  \epsilon_{\mu\nu\rho\sigma\delta\tau} M^{[\sigma\delta} \cdot P^{\tau]} \qquad B_{\mu\nu\rho} =\frac{1}{3!}  \epsilon_{\mu\nu\rho\sigma\delta\tau} M^{[\sigma \delta} \cdot K^{\tau]}
\ee
we find that they vanish identically for the minrep given above
\be
A_{\mu\nu\rho} = 0 \qquad B_{\mu\nu\rho} = 0
\ee

Computing the products of the generators of the above minimal unitary  realization corresponding to the Young tableaux $\parbox{10pt}{\YoungAAAA}$ and $\parbox{20pt}{\YoungB}$ one finds that  they vanish identically and the resulting enveloping algebra contains only the operators whose Young tableaux have two rows:

$$
\underbrace{\begin{picture}(100,50)(01.8,-18)
\put(0,5){\line(0,1){20}}
\put(00,5){\line(1,0){100}}
\put(00,15){\line(1,0){100}}
\put(00,25){\line(1,0){100}}
\put(10,5){\line(0,1){20}}
\put(20,5){\line(0,1){20}}
\put(30,5){\line(0,1){20}}
\put(40,5){\line(0,1){20}}
\put(70,5){\line(0,1){20}}
\put(80,5){\line(0,1){20}}
\put(90,5){\line(0,1){20}}
\put(100,5){\line(0,1){20}}
\put(55,10){\makebox(0,0){$\cdots$}}
\put(55,20){\makebox(0,0){$\cdots$}}
\end{picture}}_{\mbox{$n$ boxes}}
$$

\subsection{Deformations of the minimal unitary representation of SO(6,2) and the  Joseph ideal}
\label{sect-joseph-deformations}
As we saw above the generator $J_{ABCD}$ of the Joseph ideal for $SO^*(8)$ vanishes identically as an operator for the minrep obtained by quasiconformal techniques. However when one substitutes the generators of the deformed minreps one finds that $J_{ABCD}$ does not vanish identically.  However as we will show the generators of the deformed minreps satisfy certain  quadratic identities which correspond to deformations of the Joseph ideal.  

To exhibit the quadratic identities corresponding to  deformations of the Josep ideal we decompose the 
components of $J_{ABCD}$ ( $A,B,..=0,1,2,...,7$) in terms of Lorentz ($SU^*(4)$ ) covariant indices. We find that 
the totally antisymmetric tensors $A_{\mu\nu\rho}$ and $B_{\mu\nu\rho}$  ( $\mu , \nu, ..=0,1,..,5$) defined in the previous section that vanished identically for the minrep do not vanish  for the deformed minreps.  Remarkably they become self-dual and anti-self-dual tensorial operators, respectively:
\bea
A_{\mu\nu\rho} \eq \widetilde{A}_{\mu\nu\rho} \nn
B_{\mu\nu\rho} \eq -\widetilde{B}_{\mu\nu\rho} 
\label{so62def2}
\eea
The identities (\ref{mident6}) and (\ref{dmident6}) no longer hold separately but they combine and the following identity holds true for deformed generators:
\be
M_{\mu\nu} \cdot M_{\rho\sigma} + M_{\mu\sigma}\cdot M_{\nu\rho} + M_{\mu\rho}\cdot M_{\sigma\nu}   = \epsilon_{\mu\nu\rho\sigma}^{\quad\,\,\,\,\, \delta\tau} (P_{[\delta} \cdot K_{\tau]} + M_{\delta\tau} \cdot \Delta)
\label{so62def1}
\ee

The deformation of the identity (\ref{mpident6}) is as follows:
\be
\eta^{\mu\nu} M_{\mu\rho} \cdot M_{\nu\sigma} - P_{(\rho} \cdot K_{\sigma)} + 4 \eta_{\rho\sigma} = 2\mathcal{G}^2 \eta_{\rho\sigma}
\ee 
where $\mathcal{G}^2$ is quadratic Casimir of $\mathfrak{su}(2)_G$ defined in equation \ref{su2g} and involves only the deformation fermions. Note that the quadratic Casimir operator $\mathcal{G}^2$ of $\mathfrak{su}(2)_G$ is related to the quadratic Casimir operator of the deformed minrep of $SO^*(8)$ as follows \cite{Fernando:2010dp}:
\be
C_2 \left[ \mathfrak{so}^*(8) \right]_{\text{deformed}} = 8 \lb 2 -  \mathcal{G}^2 \rb
\ee
The eigenvalues $t(t+1)$ ($t=0,1/2,1,3/2,...$ ) of $\mathcal{G}^2$ label the deformations of the minrep of $SO^*(8)$. 
This is to be contrasted with the deformations of the minrep of $4d$ conformal group $SO(4,2)$ which are labelled by a continuous parameter that enters the quadratic identities  explicitly in the form of continuous helicity  \cite{Govil:2013uta}.  Since the minrep and its deformations obtained by quasiconformal methods correspond to massless conformal fields we do not expect any continuous deformations in six dimensions  since the little group of massless particles is $SO(4)=SU(2)_T\times SU(2)_A$ whose unitary representations are discretely labelled $(j_T,j_A)$ where $j_A$ and $j_T$  are non-negative integers or half integers. Furthermore for conformally massless representations either $j_T$ or $j_A$ (or both) vanishes. However we do not have a proof that continuous deformations do not exist.

For the discrete deformations of the minrep the operators corresponding to the symmetric tensor with Young Tableau
$ \parbox{20pt}{\YoungB} $ appearing in the symmetric product of the generators as shown in \ref{ytdecomp}
still vanish
On the other hand the operators with the Young tableau
\[ \parbox{10pt}
{\YoungAAAA} \]
do not vanish. These operators satisfy an eight dimensional self-duality condition which corresponds to a three form gauge field with a self-dual field strength. In $AdS_7$ they correspond to three form gauge fields that satisfy odd dimensional self-duality condition just like the three form field that descends from eleven dimensional supergravity on $AdS_7\times S^4$. In six dimensions  they correspond to conformal two form fields with a self-dual field strength, which is simply the tensor field that appears in the $(2,0)$ conformal supermultiplet whose interacting theory is believed to be dual to M-theory over $AdS_7\times S^4$. 

The symmetric tensor products of the generators of  the discrete deformations  of $SO(6,2)$ leads to a $AdS_7/CFT_6$ higher spin  algebra whose generators include Young tableaux of the form

$$\underbrace
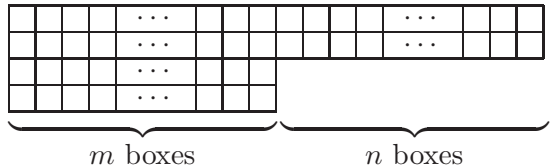
{\begin{picture}(100,30)(0,-18)
\put(0,5){\line(1,0){100}}
\put(0,15){\line(1,0){100}}
\put(0,25){\line(1,0){100}}
\put(0,-5){\line(1,0){100}}
\put(0,-15){\line(1,0){100}}
\put(0,-15){\line(0,1){40}}
\put(10,-15){\line(0,1){40}}
\put(20,-15){\line(0,1){40}}
\put(30,-15){\line(0,1){40}}
\put(40,-15){\line(0,1){40}}
\put(70,-15){\line(0,1){40}}
\put(80,-15){\line(0,1){40}}
\put(90,-15){\line(0,1){40}}
\put(100,-15){\line(0,1){40}}
\put(55,10){\makebox(0,0){$\cdots$}}
\put(55,20){\makebox(0,0){$\cdots$}}
\put(55,0){\makebox(0,0){$\cdots$}}
\put(55,-10){\makebox(0,0){$\cdots$}}
\end{picture}}_{\mbox{$m$ boxes}}
\underbrace{\begin{picture}(100,50)(01.8,-18)
\put(00,5){\line(1,0){100}}
\put(00,15){\line(1,0){100}}
\put(00,25){\line(1,0){100}}
\put(10,5){\line(0,1){20}}
\put(20,5){\line(0,1){20}}
\put(30,5){\line(0,1){20}}
\put(40,5){\line(0,1){20}}
\put(70,5){\line(0,1){20}}
\put(80,5){\line(0,1){20}}
\put(90,5){\line(0,1){20}}
\put(100,5){\line(0,1){20}}
\put(55,10){\makebox(0,0){$\cdots$}}
\put(55,20){\makebox(0,0){$\cdots$}}
\end{picture}}_{\mbox{$n$ boxes}}
$$
This suggests that the  theories based on discrete deformations of the minrep describe higher spin theories of Fradkin-Vasiliev type in $AdS_7$ coupled to tensor fields that satisfy self-duality conditions and their higher extensions corresponding to the Young tableaux \\

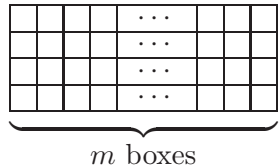
$$\underbrace{\begin{picture}(100,30)(0,-18)
\put(0,5){\line(1,0){100}}
\put(0,15){\line(1,0){100}}
\put(0,25){\line(1,0){100}}
\put(0,-5){\line(1,0){100}}
\put(0,-15){\line(1,0){100}}
\put(0,-15){\line(0,1){40}}
\put(10,-15){\line(0,1){40}}
\put(20,-15){\line(0,1){40}}
\put(30,-15){\line(0,1){40}}
\put(40,-15){\line(0,1){40}}
\put(70,-15){\line(0,1){40}}
\put(80,-15){\line(0,1){40}}
\put(90,-15){\line(0,1){40}}
\put(100,-15){\line(0,1){40}}
\put(55,10){\makebox(0,0){$\cdots$}}
\put(55,20){\makebox(0,0){$\cdots$}}
\put(55,0){\makebox(0,0){$\cdots$}}
\put(55,-10){\makebox(0,0){$\cdots$}}
\end{picture}}_{\mbox{$m$ boxes}}
$$
The study of higher spin theories based on discrete deformations of the minrep that extend Fradkin-Vasiliev type higher spin theories will be the subject of a separate study. 

\subsection{ Comparison with the covariant twistorial oscillator realization }
 Substituting the generators   $M_{AB}$ of $SO^*(8)$ realized as bilinears of covariant  twistorial oscillators (doubletons) (section \ref{sect-twist}) in equation \ref{joseph-6d} to compute the generator of the Joseph ideal one  finds that it does not vanish identically. However the non-vanishing components of the generator $J_{ABCD}$ of the Joseph ideal factorize
in a similar fashion as in the doubleton realization of $SO(4,2)$ \cite{Govil:2013uta}. 
Symbolically this factorization takes the form
\be
J_{ABCD} = (...) B_a
\ee
where $B_a$ ($a=1,2,3$) is a generator of an $SU(2)_B$ algebra that commutes with $SO^*(8)$. In terms of the covariant twistorial oscillators the generators of $SU(2)_B$ are  
\be
B_- = d^i c_i, \qquad B_+ = c^i d_i, \qquad B_0 = \frac{1}{2} ( c^i c_i - d^id_i ) 
\ee
with the quadratic Casimir
\be
\mathcal{B}^2 = B_0^2 +\frac{1}{2} ( B_+ B_- + B_- B_+ )
\ee
Acting on the subspace of the Fock space of covariant twistorial oscillators that is $SU(2)_B$ singlet the generator $J_{ABCD}$ vanishes. This subspace corresponds to the  true minrep of $SO^*(8)$ and describes a conformal scalar field in six dimensions. In fact the authors of \cite{Sezgin:2001ij} studied a purely bosonic higher spin algebra in $AdS_7$ using the doubletonic realization of $SO^*(8)$. After imposing an infinite set of constraints, restricting to an $SU(2)$ singlet sector  and modding out by an infinite ideal containing all the traces they obtain an higher spin algebra. They also state that their results can not be extended to $SU(2)$ non-singlet sectors. 

The algebra  $\mathfrak{su}(2)_B$ for the doubleton representations is the analog of  $\mathfrak{su}(2)_G$ for the deformed minreps studied above. 
The Casimir operator $\mathcal{B}^2$ of $SU(2)_B$ is related to the quadratic Casimir $C_2$ of the doubletonic realization  of $SO^*(8)$ in terms of covariant twistorial oscillators :
\be
C_2 \left[ \mathfrak{so}^*(8) \right]_{\text{doubleton}} = 8(2 -  \mathcal{B}^2)
\ee
which reflects the fact that $\mathfrak{su}(2)_B$ and $\mathfrak{so}^*(8)_{\text{doubleton}}$ form a reductive dual pair inside  $\mathfrak{sp}(16,\mathbb{R})$. However this is not the case with the deformed minreps  since $\mathfrak{su}(2)_G$ does not commute with the generators of $\mathfrak{so}^*(8)_{\text{deformed}}$.  Another critical difference is the fact that the possible eigenvalues $b(b+1)$ of $SU(2)_B$ span the entire set of irreps, i.e $b=0,1/2,1,3/2,...$. On the other hand, for a given number $P$ pairs of deformation Fermions possible eigenvalues $j(j+1)$ of $SU(2)_G $ is $ j=0,1/2,1,..,P/2 $

For the doubleton realization the quadratic identities satisfied by the generators satisfy  formally the same identities as the deformed minrep given in the previous subsection with $\mathcal{G}^2$ replaced by $\mathcal{B}^2$: 
\be
\eta^{\mu\nu} M_{\mu\rho} \cdot M_{\nu\sigma} - P_{(\rho} \cdot K_{\sigma)} + 4 \eta_{\rho\sigma} = 2\mathcal{B}^2 \eta_{\rho\sigma}
\ee 
The Casimir invariants for $SO(6,2)$ in the doubleton representation are as follows:
\bea
\label{cas1}
C_2 \eq   8 \lb 2-\mathcal{B}^2\rb \\
C_4 \eq   \frac{C_2^2}{8} - 9 C_2 \\
C'_4 \eq   96 C_2 - 6 C_2^2 \\
C_6 \eq  \frac{C_2^3}{64} - \frac{27}{8} C_2^2 + 81 C_2 
\eea

\subsection{ $AdS_7/CFT_6$ Higher spin algebras and superalgebras and their deformations}
\label{sect-hsa}
Following \cite{Eastwood:2002su,Govil:2013uta}, we will use the following definition for the standart  higher spin algebra in six dimensions:
\be
HS(6,2) = \frac{\mathscr{U}(\mathfrak{so}(6,2))}{\mathscr{J}(\mathfrak{so}(6,2))} 
\ee
where $\mathscr{U}(SO(6,2))$ is the universal enveloping algebra and $\mathscr{J}(\mathfrak{so}(6,2))$ denotes the Joseph ideal of $\mathfrak{so}(6,2)$.  The Joseph ideal vanishes identically  for the quasiconformal realization of the minrep. Therefore to construct $HS(6,2)$  one needs simply take the enveloping algebra of the minrep in the quasiconformal construction. Since the minrep of $\mathfrak{so}(6,2)$ admits deformations we define deformed $AdS_7/CFT_6$ higher spin algebras $HS(6,2;t)$ as the enveloping  algebras of $\mathfrak{so}(6,2)$ quotiented by the deformed Joseph ideal $\mathscr{J}_t(\mathfrak{so}(6,2))$ 
\be
HS(6,2;t) =\frac{\mathscr{U}(\mathfrak{so}(6,2))}{\mathscr{J}_t(\mathfrak{so}(6,2))} 
\ee
For these deformed high spin algebras  the corresponding deformed Joseph ideal vanishes identically as operator as we showed explicitly above for the conformal group in six dimensions. Deformed minreps describe massless conformal fields of higher spin labeled by the spin $t$ of the $SU(2)_G
$ subgroup which is the analogue of  helicity in $d=4$.

We saw in section \ref{sect-qc-def} that deformations of the minrep are driven by fermionic oscillators. For $P$ pairs of deformation fermions  the Fock space decomposes as the direct sum of the two spinor representations of $SO(4P)$ generated by all the bilinears of the oscillators. The centralizer of $SU(2)_G$ inside $SO(4P)$ is $USp(2P)$. Under $USp(2P)\times SU(2)_G$ the Fermionic Fock space decomposes as
\bea
2^{2P} = \sum_{r=0}^{P} (R^{r}, t=(P-r)/2)  
\eea
where $R^{r}$ is the symplectic traceless tensor of rank $r$ of $USp(2P)$ and $t$ is the spin of $SU(2)_G$.
The $USp(2P)$ invariant (singlet) subspace transforms in the spin $t=P/2$ representation of $SU(2)_G$. Therefore restricting to this invariant subspace we get a deformed minrep corresponding to a $6d$ massless conformal field transforming  as a totally symmetric tensor of rank $P$ in the spinor indices with respect to the Lorentz group $SU^*(4)$. This way one can construct all conformally massless representations of $SO(6,2)$ as deformations of the minrep by choosing $P=0,1,2,...$. The enveloping algebras of these deformed irreducible irreps define then a discrete infinity of higher spin algebras labelled by $t=P/2$. Equivalently one can simply subsitute $(P+1)\times(P+1)$ irreducible representation matrices   for generators of $SU(2)_G$ in place of the  bilinears of deformation fermions. The latter is useful for writing down the irreducible  infinite higher spin algebra without reference to its action on a representation space. 

We shall define  the $2N$ extended $AdS_7/CFT_6$ higher spin  superalgebra as the enveloping algebra of the minimal unitary realization of the super algebra $OSp(8^*|2N)$ obtained via the quasiconformal  approach.
For this algebra there are only $2N$ supersymmetry fermions and the minimal  supermultiplet of $OSp(8^*|2N)$ consists of the following massless conformal fields:
\be
\Phi^{[A_1A_2...A_N]|} \oplus \Psi_{\alpha}^{[A_1 A_2...A_{N-1}]|} \oplus \Phi_{(\alpha\beta)}^{[A_1A_2...A_{N-2}]|} \oplus  \cdots 
\ee
where $\alpha, \beta,..$ are the spinor indices of the $6d$ Lorentz group $SU^*(4)$ and $A_i$ denote the $USp(2N)$ indices.  The  generator $J_{ABCD}$ of the Joseph ideal  vanishes when acting  on the conformal scalars that are part of the minimal unitary supermultiplet. On the other fields of the minimal unitary supermultiplet   deformed quadratic identities \ref{sect-joseph-deformations} involving supersymmety fermions are satisfied. 

Deformed $AdS_7/CFT_6$ higher spin superalgebras $HS(6,2|2N;t)$ algebras are defined simply as enveloping algebras of  the deformed minimal unitary realizations of the super algebras  $OSp(8^*|2N)$ with the even subalgebra $SO^*(8)\oplus USp(2N)$ involving deformation fermions \ref{sect-qcgsusy}. As explained above restricting to $USp(2P)$ invariant sector of the Fock space of $2P$ deformation fermions one gets  a deformed minimal unitary realization of $OSp(8^*|2N)_t$ for $t =P/2$. Equivalently one can simply substitute the $(P+1)\times (P+1)$ representation matrices of $SU(2)_G$ in place of the bilinears of deformation fermions. Their enveloping algebras define a discrete infinite family of higher spin supealgebras labeled by $t=0,1/2,1,3/2,2,...$. 

\section{Discussion} \label{sec-discussion} One of our main results in this paper is the reformulation of  the minimal unitary representation of $SO^*(8)$ and its deformations in  terms of deformed twistors that transform nonlinearly under the Lorentz group in six dimensions. Their enveloping algebras lead to a discrete infinite family of $AdS_7/CFT_6$ higher spin algebras labelled by the spin of an $SU(2)$ symmetry for which certain deformations of the Joseph ideal vanish. Remarkably these deformations involve (anti-)self-duality of the $6d$ tensorial operator which is the analog of Pauli-Lubanski vector in four dimensions. These results carry  to superalgebras  $OSp(8^*|2N)$ and one finds a discrete infinity of $AdS_7/CFT_6$ higher spin superalgebras.  As we argued in our previous work \cite{Govil:2013uta}
for $AdS_5/CFT_4$ algebras our results suggest the existence of a family of (supersymmetric) higher spin theories in $AdS_7$ that are dual to free (super) CFT's or to interacting but integrable (supersymmetric) CFT's in six dimensions. This is suggested by the fact that, in contrast to $AdS_4/CFT_3$ higher spin algebras the higher dimensional algebras are realized in terms of deformed twistors that transform nonlinearly and by the results of \cite{Govil:2012rh} on the precise mapping between the deformed minreps of $D(2,1;\alpha)$  and the spectra of certain integrable supersymmetrical quantum mechanical models. 
Of particular interest are the higher spin superalgebras based on $OSp(8^*|4)$ and $OSp(8^*|8)$ whose minimal unitary supermultiplets reduce to $N=4$ Yang-Mills supermultiplet and $N=8$ supergravity multiplet under dimensional reduction to four dimensions\cite{Chiodaroli:2011pp}.  The minimal unitary supermultiplet of $OSp(8^*|4)$ is the $6d$ (2,0) conformal tensor multiplet \cite{Gunaydin:1984wc,Gunaydin:1999ci} whose interacting theory is believed to be dual to M-theory over $AdS_7\times S^4$\cite{Maldacena:1997re}. Whether there exists a limit of this interacting theory that is dual to a higher spin theory in $AdS_7$ is an interesting open problem. Our results in section 3.2 suggest  that such a limit should exist. On the other hand it is not known if there exists an interacting non-metric $(4,0)$  supergravity theory based on the minimal unitary supermultiplet of $OSp(8^*|8)$  \cite{Hull:2000rr,Chiodaroli:2011pp}.

{\bf Acknowledgements:} Main results of this paper were announced in GGI Workshop on Higher spin symmetries (May 6-9, 2013)  and Summer Institute (Aug. 2013)  at ENS in Paris by MG and at Helmholtz International Summer School (Sept. 2013) at Dubna by KG. We would like to thank the organizers of these workshops and institutes  for their kind hospitality where part of this work was carried out. We enjoyed discussions with many of their participants. We are especially grateful  to Misha Vasiliev and Eugene Skvortsov for stimulating discussions regarding higher spin theories and Dmytro Volin regarding quasiconformal realizations. This research was supported in part by the US National Science Foundation under grants PHY-1213183, PHY-08-55356 and DOE Grant No: DE-SC0010534. One of us (MG) would like to thank the Theory Division at CERN and the Albert Einstein Institute, Potsdam, for their kind hospitality  where this work was completed.


\begin{appendices}
\section{Clifford algebra conventions for doubleton realization}
\label{sigmaold}
In this appendix, we will give the $6d$ analogs of $\sigma_\mu$ matrices (in mostly positive metric) used in section \ref{sect-twist} as defined in \cite{Gunaydin:1999ci} for the doubleton representations of $SO^*(8)$. Pauli matrices
\begin{equation}
\sigma_0 = \begin{pmatrix}
1 & 0 \\
0 & 1
\end{pmatrix},
\;\; \sigma_1 = \begin{pmatrix}
0 & 1 \\
1 & 0
\end{pmatrix},
\;\; \sigma_2 = \begin{pmatrix}
0 & -i \\
i & 0
\end{pmatrix},
\;\; \sigma_3 = \begin{pmatrix}
1 & 0 \\
0 & -1
\end{pmatrix}.
\;\;
\end{equation}
satisfy 
\begin{equation}
\sigma^\mu \bar \sigma^\nu + \sigma^\nu \bar \sigma^\mu = 2 \eta^{\mu \nu}.
\end{equation}
Their $6d$ counterparts are defined as \cite{Gunaydin:1999ci}
\begin{subequations}
\begin{align}
\Sigma^0 &= -i \sigma_2 \otimes \sigma_3 & \bar \Sigma^0 &= -i \sigma_2 \otimes \sigma_3 \\
\Sigma^1 &= i \sigma_2 \otimes \sigma_0 & \bar \Sigma^1 &= -i \sigma_2 \otimes \sigma_0 \\
\Sigma^2 &= i \sigma_1 \otimes \sigma_2 & \bar \Sigma^2 &= -i \sigma_1 \otimes \sigma_2 \\
\Sigma^3 &= i \sigma_3 \otimes \sigma_2 & \bar \Sigma^3 &= -i \sigma_3 \otimes \sigma_2 \\
\Sigma^4 &= \sigma_0 \otimes \sigma_2 & \bar \Sigma^4 &= \sigma_0 \otimes \sigma_2 \\
\Sigma^5 &= \sigma_2 \otimes \sigma_1 & \bar \Sigma^5 &= \sigma_2 \otimes \sigma_1.
\end{align}
\end{subequations}
with the convention that the six dimensional $\Sigma^\mu$ have lower spinorial indices while the $\bar \Sigma^\mu$ have upper spinorial indices. 

\section{Clifford algebra conventions for deformed twistors}
\label{app-conventions}
In order to make contact with the spinor helicity literature in 6d, we will use the mostly positive metric and follow the conventions of \cite{Cheung:2009dc} for 6d analogs of Pauli matrices in the formulation of deformed minreps in terms of deformed twistors. We use a hat over the 6d sigma matrices in order to avoid confusion with the standard Pauli matrices. 
\begin{subequations}
\begin{align}
\hat{\sigma}^0 &= i \sigma_1 \otimes \sigma_2 & \bar{\hat{\sigma}}^0 &= -i \sigma_1 \otimes \sigma_2 \\
\hat{\sigma}^1 &= i \sigma_2 \otimes \sigma_3 & \bar{\hat{ \sigma}}^1 &= i \sigma_2 \otimes \sigma_3 \\
\hat{\sigma}^2 &= - \sigma_2 \otimes \sigma_0 & \bar{\hat{\sigma}}^2 &= \sigma_2 \otimes \sigma_0 \\
\hat{\sigma}^3 &= -i \sigma_2 \otimes \sigma_1 & \bar{\hat{\sigma}}^3 &= -i \sigma_2 \otimes \sigma_1 \\
\hat{\sigma}^4 &= - \sigma_3 \otimes \sigma_2 & \bar{\hat{\sigma}}^4 &= \sigma_3 \otimes \sigma_2 \\
\hat{\sigma}^5 &= i \sigma_0 \otimes \sigma_2 & \bar{\hat{\sigma}}^5 &= i \sigma_0 \otimes \sigma_2.
\end{align}
\end{subequations}
They satisfy :
\begin{equation}
\hat{\sigma}^\mu \bar{\hat{\sigma}}^\nu + \hat{\sigma}^\nu \bar{\hat{\sigma}}^\mu = -2 \eta^{\mu \nu}.
\end{equation}
Again we adopt the convention that the six dimensional $\hat{\sigma}^\mu$ have lower spinorial indices while the $\bar{\hat{\sigma}}^\mu$ have upper spinorial indices. 
With these conventions, we define:
\be
P_{\a\b}=(\hat{\sigma}^\mu P_\mu)_{\a\b} = \begin{pmatrix}
0                   & iP_4 + P_5 & P_1 + iP_2 & P_0 - P_3 \\
-iP_4 - P_5 & 0                   & -P_0 - P_3 & -P_1 + iP_2 \\
-P_1 - iP_2 & P_0 + P_3  & 0                  & -iP_4 + P_5 \\
-P_0 + P_3 & P_1 - iP_2  & iP_4 - P_5 &  0 
\end{pmatrix}
\ee

\be
K^{\a\b}=(\bar{\hat{\sigma}}^\mu K_\mu)^{\a\b} = \begin{pmatrix}
0                   & -iK_4 + K_5 & K_1 - iK_2 & -K_0 - K_3 \\
iK_4 - K_5 & 0                   & K_0 - K_3 & -K_1 - iK_2 \\
-K_1 + iK_2 & -K_0 + K_3  & 0                  & iK_4 + K_5 \\
K_0 + K_3 & K_1 + iK_2  & -iK_4 - K_5 &  0 
\end{pmatrix}
\ee

\section{The quasiconformal realization of the minimal unitary representation of $SO(6,2)$ in compact 3-grading}
\label{app-qcg3grading}
In this appendix we provide the formulas for the quasiconformal realization of generators of $SO(6,2)$ and their deformations  in compact 3-grading following \cite{Fernando:2010dp}. We shall give the formulas with the deformation fermions included. The generators for the true minrep can be obtained simply by setting the deformation fermions to zero.

Consider the compact three graded decomposition of the Lie algebra of $SO^*(8)$ determined by the conformal Hamiltonian $H$
\be
\mathfrak{so}^*(8) = \mathfrak{C}^- \oplus \mathfrak{C}^0 \oplus \mathfrak{C}^+
\ee
where $\mathfrak{C}^0 = \mathfrak{su}(4) \oplus \mathfrak{u}(1)$. We shall label the generators in $\mathfrak{C}^\pm$ and $\mathfrak{C}^0$  as follows:
\bea
(W_m,X_m,N_-,B_-) &\in& \mathfrak{C}^- \\
(D_m,E_m,D^m,E_m,T_{\pm,0},A_{\pm,0},J,H) &\in& \mathfrak{C}^0 \\
(W^m,X^m,N_-,B_+) &\in& \mathfrak{C}^+
\eea
where $m,n,..=1,2$. 
The generators of $\mathfrak{su}(4)$ algebra in $\mathfrak{C}^0$ subspace  has a 3-graded decomposition with respect to its  $  \mathfrak{su}(2)_T \oplus \mathfrak{su}(2)_A \oplus \mathfrak{u}(1)_J$ subalgebra where the $U(1)_J$ generator  $J$ determines the 3-grading of $\mathfrak{su}(4)$. The generators for $\mathfrak{su}(2)_T$ were given in equation \ref{gen-su(2)T} (for the true minrep without deformation fermions $\mathfrak{su}(2)_T$ reduces simply to  $\mathfrak{su}(2)_S$ whose generators were given in equation \ref{gen-su(2)S}) and those of $\mathfrak{su}(2)_A$ are as follows:
\be
A_+ = a^1 a_2 + b^1 b_2, \qquad A_- = a_1 a^2 + b_1 b^2, \qquad A_0 = \half \lb a^1a_1 - a^2a_2 + b^1b_1 - b^2b_2 \rb \nonumber
\label{gen-su(2)A}
\ee
and they satisfy:
\be
\comm{A_+}{A_-} = 2A_0, \qquad \comm{A_0}{A_\pm} = \pm A_\pm
\ee
The generators $D_m.E_m,D^m,E^m$ belonging to the coset $SU(4)/SU(2)\times SU(2)\times U(1)$ are realized as bilinears of the oscillators  $a_m,b_m$  and  the following ``singular" oscillators :
\be
A_{\ml_\pm} = \frac{1}{\sqrt{2}} \lb x+ip - \frac{\ml_\pm}{x} \rb, \qquad A_{\mk_\pm} = \frac{1}{\sqrt{2}} \lb x+ip - \frac{\mk_\pm}{x} \rb 
\label{sing-osc}
\ee
where
\be
\ml_\pm = 2 \lb T_0 \pm T_- -\frac{3}{4} \rb, \qquad \mk_\pm = -2 \lb T_0 \pm T_+ +\frac{3}{4} \rb 
\ee
They  satisfy the  commutation relations:
\begin{align}
\comm{\ml_+}{\ml_-} &= 2(\ml_+ - \ml_-), \qquad \comm{\mk_+}{\mk_-} = 2(\mk_+ - \mk_-) \\
\comm{\ml_\pm}{\mk_\pm} &= 2(\ml_\mp - \mk_\mp), \qquad \comm{\ml_\pm}{\mk_\mp} = -2(\ml_\pm - \mk_\mp) 
\label{lk-comm}
\end{align}
In general for two singular oscillators defined in terms of operators  $\ml_1$ and $\ml_2$that
commute with the singlet coordinate $x$  but not with each other we have we have 
\bea
\comm{A_{\ml_1}}{A_{\ml_2}} \eq \frac{1}{2} \lb \frac{\ml_2-\ml_1}{x^2} + \frac{\comm{\ml_1}{\ml_2}}{x^2} \rb \\
\comm{A^\dagger_{\ml_1}}{A^\dagger_{\ml_2}} \eq \frac{1}{2} \lb \frac{\ml^\dagger_1-\ml^\dagger_2}{x^2} + \frac{\comm{\ml^\dagger_1}{\ml^\dagger_2}}{x^2} \rb \\
\comm{A_{\ml_1}}{A^\dagger_{\ml_2}} \eq 1 + \frac{1}{2} \lb \frac{\ml_1 + \ml^\dagger_2}{x^2} + \frac{\comm{\ml_1}{\ml^\dagger_2}}{x^2} \rb 
\eea
where $A^\dagger_{\ml_\pm} = \frac{1}{\sqrt{2}} \lb x+ip - \frac{\ml^\dagger_\pm}{x} \rb$. The commutation relations (\ref{lk-comm}) lead to the commutation relations for $A_{\ml_\pm}$ and $A_{\mk_\pm}$ as follows:
\begin{align}
\comm{A_{\ml_+}}{A_{\ml_-}} &= \frac{(\ml_+ - \ml_-)}{2x^2}, \qquad \comm{A_{\mk_+}}{A_{\mk_-}} = \frac{(\mk_+ - \mk_-)}{2x^2} \\
\comm{A_{\ml_+}}{A_{\mk_-}} &= -\frac{3(\ml_+ - \mk_-)}{2x^2}, \qquad \comm{A_{\ml_-}}{A_{\mk_+}} = -\frac{3(\ml_- - \mk_+)}{2x^2} 
\end{align}

The generator that determines the compact 3-grading of $SO^*(8)$ is given as follows:
\begin{equation}
H = H_a + H_b + H_\odot
\end{equation}
where
\be
H_{\odot} = \frac{1}{4} \lb A_{\ml_-}A^\dagger_{\ml_-} + A_{\mk_+}A^\dagger_{\mk_+} + \ml_- + \mk_+ - 1 \rb
\end{equation}
and
\begin{equation}
H_a = \frac{1}{2} \left( N_a + 2 \right),
\qquad 
H_b = \frac{1}{2} \left( N_b + 2 \right)
\ee
are simply  the Hamiltonians of standard bosonic  oscillators of $a$- and $b$-type ($N_a=a^1a_1+a^2a_2, N_b=b^1b_1+b^2b_2$). This $\mathfrak{u}(1)$ generator is the $AdS$ energy or the conformal Hamiltonian when $SO^*(8) \simeq SO(6,2)$ is taken as the seven dimensional $AdS$ group or the six dimensional conformal group, respectively.

The generator that determines the 3-grading of $\mathfrak{su}(4)$ is as follows:
\begin{equation}
J = H_a + H_b - H_\odot
\end{equation}
The $SU(4)/SU(2)_S \times SU(2)_A \times U(1)_J$  coset generators can be written as
\begin{align}
D_m &= \frac{1}{\sqrt{2}} \lb a_m A^\dag_{\ml_+}  + b_m A^\dag_{\mk_-} \rb, & D^m = \frac{1}{\sqrt{2}} \lb A_{\ml_+} a^m + A_{\mk_-}b^m \rb \\ 
E_m &= \frac{1}{\sqrt{2}} \lb  a_m A^\dag_{\ml_-}  - b_m A^\dag_{\mk_+} \rb, & E^m = \frac{1}{\sqrt{2}} \lb A_{\ml_-} a^m - A_{\mk_+}b^m  \rb
\end{align}
where $m,n=1,2$. They close into   the generators of the subgroup $ SU(2)_S \times SU(2)_A \times U(1)_J $ given in (\ref{gen-su(2)T}) and (\ref{gen-su(2)A}), which do not involve singular oscillators.
Then the $\mathfrak{su}(4)$ algebra can be rewritten in a fully $SU(2)_S\times SU(2)_A$ covariant form 
\begin{equation}
\begin{split}
&\comm{S^{m^\prime}_{~n^\prime}}{S^{k^\prime}_{~l^\prime}}
= \delta^{k^\prime}_{n^\prime} \, S^{m^\prime}_{~l^\prime}
  - \delta^{m^\prime}_{l^\prime} \, S^{k^\prime}_{~n^\prime}
\qquad \qquad \qquad
\comm{A^m_{~n}}{A^k_{~l}}
= \delta^k_n \, A^m_{~l} - \delta^m_l \, A^k_{~n}
\\
&\comm{C^{m^\prime m}}{C_{n^\prime n}}
= \delta^m_n \, S^{m^\prime}_{~n^\prime}
  + \delta^{m^\prime}_{n^\prime} \, A^m_{~n}
  + \delta^{m^\prime}_{n^\prime} \delta^m_n \, J
\\
&\comm{S^{m^\prime}_{~n^\prime}}{C^{k^\prime m}}
= \delta^{k^\prime}_{n^\prime} \, C^{m^\prime m}
  - \frac{1}{2} \delta^{m^\prime}_{n^\prime} \, C^{k^\prime m}
\qquad
\comm{A^m_{~n}}{C^{m^\prime k}}
= \delta^k_n \, C^{m^\prime m}
  - \frac{1}{2} \delta^m_n \, C^{m^\prime k} \,.
\end{split}
\end{equation}
where we have labeled the generators of $\mathfrak{su}(2)_S$ and
$\mathfrak{su}(2)_A$ as $S^{m^\prime}_{~n^\prime}$ ($m^\prime, n^\prime =1,2$) and $A^m_{~n}$, respectively:
\begin{equation}
\begin{aligned}
S^1_{~1} &= - S^2_{~2} = S_0
\\
A^1_{~1} &= - A^2_{~2} = A_0
\end{aligned}
\qquad \qquad \qquad
\begin{aligned}
S^1_{~2} &= S_+
\\
A^1_{~2} &= A_+
\end{aligned}
\qquad \qquad \qquad
\begin{aligned}
S^2_{~1} = \left( {S^1_{~2}} \right)^\dag &= S_-
\\
A^2_{~1} = \left( {A^1_{~2}} \right)^\dag &= A_-
\end{aligned}
\end{equation}
and defined 
\begin{align}
C_{1m} &= D_m + E_m, \qquad C_{2m} = D_m - E_m \\
C^{1m} &= D^m + E^m, \qquad C^{2m} = D^m - E^m 
\end{align}
The generators belonging to $\mathfrak{C}^-$ are given as follows:
\begin{align}
W_m &= \frac{1}{\sqrt{2}} \lb A_{\mk_+} a_m + A_{\ml_-}b_m \rb, & X_m = \frac{1}{\sqrt{2}} \lb A_{\mk_-} a_m - A_{\ml_+}b_m \rb \\
N_-  &= a_1 b_2 - a_2 b_1, & B_-  = \frac{1}{4} \lb A_{\mk_+}A_{-\mk_+^\dagger} + A_{\ml_-}A_{-\ml_-^\dagger} \rb
\end{align}
and the generators in $\mathfrak{C}^+$ are given by their hermitian conjugates:
\begin{align}
W^m &= \frac{1}{\sqrt{2}} \lb a^m A^\dag_{\mk_+}  + b^m A^\dag_{\ml_-} \rb, &X^m = \frac{1}{\sqrt{2}} \lb a^m A^\dag_{\mk_-}  - b^m A^\dag_{\ml_+} \rb \\
N_+ &= a^1 b^2 - a^2 b^1, & B_+  = \frac{1}{4} \lb A^\dagger_{-\mk_+^\dagger}A^\dagger_{\mk_+} + A^\dagger_{-\ml_-^\dagger}A^\dagger_{\ml_-} \rb
\end{align}

The commutators $\comm{\mathfrak{C}^-}{\mathfrak{C}^+}$ close into
$\mathfrak{C}^0$:
\begin{equation}
\begin{aligned}
\comm{W_m}{W^n} &= 2 \lb \delta^n_m \, H  + A^n_{~m} \rb + \d^n_m \lb T_- + T_ + \rb  \\
\comm{W_m}{X^n} &= \delta^n_m \, \lb 2T_0 - T_- + T_+ \rb \\ 
\comm{W_m}{N_+} &= \epsilon_{mn} \, E^{n} \\
\comm{W_m}{B_+} &= D_{m} \\
\comm{N_-}{N_+} &= H + J \end{aligned}
\qquad \qquad
\begin{aligned} 
\comm{X_m}{X^n} &= 2 \lb \delta^n_m \, H  + A^n_{~m} \rb - \d^n_m \lb T_- + T_ + \rb    \\
\comm{X_m}{W^n} &= \delta^n_m \, \lb 2T_0 + T_- - T_+ \rb \\ 
\comm{X_m}{N_+} &= \epsilon_{mn} D^{n}   \\
\comm{X_m}{B_+}  &= E_{m}   \\
\comm{B_-}{B_+}  &= H - J
\end{aligned}
\end{equation}

\section{Commutation relations of  deformed twistors}
\label{app-YZcomm}

We should note that the deformed twistorial operators transform nonlinearly under the Lorentz group. However their bilinears that enter the generators of the $SO^*(8)$ transform covariantly with respect to the 
Lorentz group $SU^*(4)$. We give below some of the commutators of deformed twistorial oscillators:
\be
\begin{split}
\comm{Z_1^{\s 1}}{Z_1^{\s 2}} & = -\frac{3}{2x} {Z_1^{\s 2}}, \\
\comm{Z_1^{\s 1}}{Z_2^{\s 1}} & = \frac{1}{2x} {Z_2^{\s 1}},  \\
\comm{Z_1^{\s 1}}{Z_2^{\s 2}} & = \frac{1}{2x} {\lb Z_1^{\s 1}-Z_2^{\s 2} \rb}, \\
\comm{Z_1^{\s 1}}{Z_3^{\s 1}} & = -\frac{1}{2x} {\lb Z_1^{\s 1}-Z_3^{\s 1} \rb}, \\
\comm{Z_1^{\s 1}}{Z_3^{\s 2}} & = -\frac{1}{2x} {\lb 2Z_1^{\s 2}+Z_3^{\s 2} \rb}, \\
\comm{Z_1^{\s 1}}{Z_4^{\s 1}} & = \frac{1}{2x} {Z_4^{\s 1}},  \\
\comm{Z_1^{\s 1}}{Z_4^{\s 2}} & = \frac{1}{2x} {\lb Z_1^{\s 1}-Z_4^{\s 2} \rb},
\end{split}
\qquad
\begin{split}
\comm{Z_1^{\s 1}}{\widetilde{Z}_1^{\s 1}} & = \frac{1}{2x} {\lb Z_1^{\s 1}-\widetilde{Z}_1^{\s 1} \rb}, \\                                          
\comm{Z_1^{\s 1}}{\widetilde{Z}_1^{\s 2}} & = -\frac{3}{2x} {\widetilde{Z}_1^{\s 2}}, \\
\comm{Z_1^{\s 1}}{\widetilde{Z}_2^{\s 1}} & = \frac{1}{2x} {\widetilde{Z}_2^{\s 1}},  \\
\comm{Z_1^{\s 1}}{\widetilde{Z}_2^{\s 2}} & = \frac{1}{2x} {\lb Z_1^{\s 1}-\widetilde{Z}_2^{\s 2} \rb}, \\
\comm{Z_1^{\s 1}}{\widetilde{Z}_3^{\s 1}} & = -\frac{1}{2x} {\lb \widetilde{Z}_1^{\s 1}-Z_3^{\s 1} \rb}, \\
\comm{Z_1^{\s 1}}{\widetilde{Z}_3^{\s 2}} & = -\frac{1}{2x} {\lb 2Z_1^{\s 2}+\widetilde{Z}_3^{\s 2} \rb}, \\
\comm{Z_1^{\s 1}}{\widetilde{Z}_4^{\s 1}} & = \frac{1}{2x} {\widetilde{Z}_4^{\s 1}},  \\
\comm{Z_1^{\s 1}}{\widetilde{Z}_4^{\s 2}} & = \frac{1}{2x} {\lb Z_1^{\s 1}-\widetilde{Z}_4^{\s 2} \rb}
\end{split}
\ee

\be
\begin{split}
\comm{Z_1^{\s 1}}{Y^{11}} & = \frac{1}{2x} {Y^{11}}, \\                                          
\comm{Z_1^{\s 1}}{Y^{12}} & = -\frac{1}{2x} {\lb Z_1^{\s 1} + Y^{12} + 2 \rb}, \\
\comm{Z_1^{\s 1}}{Y^{21}} & = -\frac{1}{2x} {\lb Z_1^{\s 1} - Y^{21} \rb},  \\
\comm{Z_1^{\s 1}}{Y^{22}} & = -\frac{1}{2x} {\lb 2Z_1^{\s 2}+Y^{22} \rb}, \\
\comm{Z_1^{\s 1}}{Y^{31}} & = \frac{1}{2x} {Y^{31}}, \\
\comm{Z_1^{\s 1}}{Y^{32}} & = \frac{1}{2x} {\lb Z_1^{\s 1}-Y^{32} \rb}, \\
\comm{Z_1^{\s 1}}{Y^{41}} & = \frac{1}{2x} {\lb Z_1^{\s 1} + Y^{41} \rb},  \\
\comm{Z_1^{\s 1}}{Y^{42}} & = \frac{1}{2x} {\lb 2Z_1^{\s 2}-Y^{42} \rb},
\end{split}
\qquad
\begin{split}
\comm{Z_1^{\s 1}}{\widetilde{Y}^{11}} & = \frac{1}{2x} {\widetilde{Y}^{11}}, \\                                          
\comm{Z_1^{\s 1}}{\widetilde{Y}^{12}} & = -\frac{1}{2x} {\lb Z_1^{\s 1} + \widetilde{Y}^{12} + 2 \rb}, \\
\comm{Z_1^{\s 1}}{\widetilde{Y}^{21}} & = -\frac{1}{2x} {\lb \widetilde{Z}_1^{\s 1} - Y^{21} \rb},  \\
\comm{Z_1^{\s 1}}{\widetilde{Y}^{22}} & = -\frac{1}{2x} {\lb 2Z_1^{\s 2}+\widetilde{Y}^{22} \rb}, \\
\comm{Z_1^{\s 1}}{\widetilde{Y}^{31}} & = \frac{1}{2x} {\widetilde{Y}^{31}}, \\
\comm{Z_1^{\s 1}}{\widetilde{Y}^{32}} & = \frac{1}{2x} {\lb Z_1^{\s 1}-\widetilde{Y}^{32} \rb}, \\
\comm{Z_1^{\s 1}}{\widetilde{Y}^{41}} & = \frac{1}{2x} {\lb \widetilde{Z}_1^{\s 1} + Y^{41} \rb},  \\
\comm{Z_1^{\s 1}}{\widetilde{Y}^{42}} & = \frac{1}{2x} {\lb 2Z_1^{\s 2}-\widetilde{Y}^{42} \rb}
\end{split}
\ee

\be
\begin{split}
\comm{Y^{11}}{Y^{12}} & = -\frac{3}{2x} {Y^{11}}, \\
\comm{Y^{11}}{Y^{21}} & = -\frac{1}{2x} {Y^{11}},  \\
\comm{Y^{11}}{Y^{22}} & = -\frac{1}{x} {\lb Y^{12}+Y^{21} \rb}, \\
\comm{Y^{11}}{Y^{31}} & = 0, \\
\comm{Y^{11}}{Y^{32}} & = \frac{1}{2x} {\lb Y^{11}-2Y^{31} \rb}, \\
\comm{Y^{11}}{Y^{41}} & = \frac{1}{2x} { Y^{11}},  \\
\comm{Y^{11}}{Y^{42}} & = \frac{1}{x} {\lb Y^{12}-Y^{41} \rb},
\end{split}
\qquad
\begin{split}
\comm{Y^{11}}{\widetilde{Y}^{11}} & = 0, \\                                          
\comm{Y^{11}}{\widetilde{Y}^{12}} & = -\frac{3}{2x} {\widetilde{Y}^{11}}, \\
\comm{Y^{11}}{\widetilde{Y}^{21}} & = -\frac{1}{2x} {\widetilde{Y}^{11}},  \\
\comm{Y^{11}}{\widetilde{Y}^{22}} & = -\frac{1}{x} {\lb \widetilde{Y}^{12}+\widetilde{Y}^{21} \rb}, \\
\comm{Y^{11}}{\widetilde{Y}^{31}} & = 0, \\
\comm{Y^{11}}{\widetilde{Y}^{32}} & = \frac{1}{2x} {\lb \widetilde{Y}^{11}-2\widetilde{Y}^{31} \rb}, \\
\comm{Y^{11}}{\widetilde{Y}^{41}} & = \frac{1}{2x} { \widetilde{Y}^{11}},  \\
\comm{Y^{11}}{\widetilde{Y}^{42}} & = \frac{1}{x} {\lb \widetilde{Y}^{12}-\widetilde{Y}^{41} \rb},
\end{split}
\ee

\end{appendices}

\providecommand{\href}[2]{#2}\begingroup\raggedright\endgroup

\end{document}